\documentclass[aps,prx,twocolumn,superscriptaddress,showpacs,floatfix,10pt]{revtex4-1}
\usepackage{amsmath, amssymb}
\usepackage{braket}
\usepackage{graphicx, subfigure}
\usepackage{color}
\usepackage{epstopdf}
\usepackage{natbib}
\usepackage{comment}

\begin{document}

\title{Standard model of the rare earths, analyzed from the Hubbard I approximation}


\author{I. L. M. Locht}
\affiliation{Dept.\ of Physics and Astronomy, Uppsala University, Box 516, SE-75120 Uppsala, Sweden}
\affiliation{Radboud University, Institute for Molecules and Materials, Heyendaalseweg 135, 6525 AJ Nijmegen, the Netherlands}
\author{Y. O. Kvashnin}
\affiliation{Dept.\ of Physics and Astronomy, Uppsala University, Box 516, SE-75120 Uppsala, Sweden}
\author{D. C. M. Rodrigues}
\affiliation{Faculdade de F\'isica, Universidade Federal do Par\'a, Bel\'em, PA, Brazil}
\affiliation{Dept.\ of Physics and Astronomy, Uppsala University, Box 516, SE-75120 Uppsala, Sweden}
\author{M. Pereiro}
\affiliation{Dept.\ of Physics and Astronomy, Uppsala University, Box 516, SE-75120 Uppsala, Sweden}
\author{A. Bergman}
\affiliation{Dept.\ of Physics and Astronomy, Uppsala University, Box 516, SE-75120 Uppsala, Sweden}
\author{L. Bergqvist}
\affiliation{Department of Materials and Nanophysics, School of Information and Communication Technology, Electrum 229, Royal Institute of Technology (KTH), SE-16440 Kista, Sweden}
\affiliation{Swedish e-Science Research Centre (SeRC), KTH Royal Institute of Technology, SE-100 44 Stockholm, Sweden}
\author{A. I. Lichtenstein}
\affiliation{Hamburg University, Inst Theoret Phys, D-20355 Hamburg, Germany}
\author{M. I. Katsnelson}
\affiliation{Radboud University, Institute for Molecules and Materials, Heyendaalseweg 135, 6525 AJ Nijmegen, the Netherlands}
\author{A. Delin}
\affiliation{Dept.\ of Physics and Astronomy, Uppsala University, Box 516, SE-75120 Uppsala, Sweden}
\affiliation{Swedish e-Science Research Centre (SeRC), KTH Royal Institute of Technology, SE-100 44 Stockholm, Sweden}
\affiliation{Department of Materials and Nanophysics, School of Information and Communication Technology, Electrum 229, Royal Institute of Technology (KTH), SE-16440 Kista, Sweden}
\author{A. B. Klautau}
\affiliation{Faculdade de F\'isica, Universidade Federal do Par\'a, Bel\'em, PA, Brazil}
\author{B. Johansson}
\affiliation{Dept.\ of Physics and Astronomy, Uppsala University, Box 516, SE-75120 Uppsala, Sweden}
\affiliation{Department of Material Science and Engineering, KTH-Royal Institute of Technology, Stockholm SE-10044, Sweden}
\author{I. Di Marco}
\affiliation{Dept.\ of Physics and Astronomy, Uppsala University, Box 516, SE-75120 Uppsala, Sweden}
\author{O. Eriksson}
\affiliation{Dept.\ of Physics and Astronomy, Uppsala University, Box 516, SE-75120 Uppsala, Sweden}

\date{\today}


\begin{abstract}
In this work we examine critically the electronic structure of the rare-earth elements by use of the so-called Hubbard I approximation. From the theoretical side all measured features of both occupied and unoccupied states are reproduced, without significant deviations between observations and theory. We also examine cohesive properties like the equilibrium volume and bulk modulus, where we find, in general, a good agreement between theory and measurements. In addition we have reproduced the spin and orbital moments of these elements, as they are reflected from measurements of the saturation moment. We have also employed the Hubbard I approximation to extract the interatomic exchange parameters of an effective spin Hamiltonian for the heavy rare earths.  We show that the Hubbard I approximation gives results which are consistent with calculations where $4f$ electrons are treated as core states for Gd. The latter approach was also used to address the series of the heavy/late rare-earths. Via Monte Carlo simulations we obtained ordering temperatures which reproduce measurements within about $20\%$. We have further illustrated the accuracy of these exchange parameters by comparing measured and calculated magnetic configurations for the heavy rare earths and the magnon dispersion for Gd. The Hubbard I approximation is compared to other theories of the electronic structure, and we argue that it is superior. We discuss the relevance of our results in general, and how this makes it possible to treat the electronic structure of materials containing rare-earth elements, such as permanent magnets, magnetostrictive compounds, photovoltaics, optical fibers, topological insulators, and molecular magnets.

\end{abstract}

\pacs{71.20.Eh, 75.30.-m, 79.60.-i, 61.50.Ah}
\maketitle




\section{Introduction}
Rare-earth (RE) elements are found in a wide range of functional materials, with diverse applications. Prominent examples may be found in the Nd$_2$Fe$_{14}$B compound with its outstanding properties as a permanent magnet~\cite{croat}, and the so called Terfenol materials, Tb$_x$Dy$_{1-x}$Fe$_2$, that have excellent magnetostrictive properties~\cite{terfenol1,terfenol2,terfenol3,terfenol4} and are used in e.g. applications involving transducers. Multiferroism, the coupling between ferroelectric and ferromagnetic degrees of freedom often involves rare-earth based compounds, an example is TbMnO$_3$~\cite{multiferroics}. Furthermore, rare-earth impurities in semiconductors are frequently used in optical fibers for signal enhancement~\cite{fibers}, in high-power fiber lasers~\cite{lasers}, as well as in photovoltaic applications~\cite{photovolt}. Examples can also be found of rare-earth containing compounds with topologically protected surface states, e.g. in the mixed valence semiconductor SmB$_6$~\cite{smb6}. Moreover, there is recent interest in rare-earth elements in molecular magnets, e.g. in so called  phthalocyanine double-deckers~\cite{doubledeck}. The examples mentioned above all represent wide research fields with different focus, ranging from applied technology to more fundamental aspects of the electronic structure of solids.

In all the materials listed above, the $4f$ shell of the rare-earth element plays an active role in determining the materials unique properties. Hence, it is of fundamental importance to establish a sound and practical theory of this $4f$ shell, not only for the materials listed above, but for the entire class of rare-earth based compounds. The standard model of the rare-earths does indeed provide such a model~\cite{jensmac,handbook}, and it has been extremely successful in describing the properties of the rare-earth elements and their compounds. The standard model of the rare-earths identifies the $4f$ shell as atomic-like, with vanishingly small wave-function overlap between neighboring atoms of the crystal lattice (the only exception here is Ce which sometimes forms a narrow band~\cite{johansson}), with the consequence that band-dispersion is absent. Then the main interaction of the $4f$ shell with the lattice is a crystal field splitting of a Russel-Saunders coupled J-manifold, and the establishment of the valency (most isolated rare-earths atoms are divalent whereas most elements are trivalent in the condensed phase). Interesting intermediate valence states are well-known in compounds like e.g. SmS, TmTe and Yb under pressure~\cite{jensmac,handbook}. The rare-earths display such robust behavior of the physical and chemical properties, both amongst elements and compounds, that the standard model is expected to apply equally well for them all. This fortunate situation allows for testing electronic structure theories for subgroups of rare-earth systems, while allowing for conclusions of the applicability of any theory over wide classes of rare-earth based materials.

Even though the standard model is well established in analyzing experimental data of rare-earth based materials~\cite{PhysRevB.11.2836, Johansson198449, PhysRevB.20.1315}, ab-initio based electronic structure theory has struggled in making a solid connection to it. It is well known that typical parametrizations of the exchange-correlation potential of density functional theory~\cite{dft1,dft2} are inadequate. For this reason theories like LDA+U~\cite{zaanen}, self-interaction correction (SIC)~\cite{sic}, orbital polarization~\cite{op}, and a treatment of the $4f$ shell as core-like~\cite{skriver}, have all been explored. Each of these theories describe some properties of the material at hand, sometimes with exceptional accuracy~\cite{skriver}, but a complete set of relevant properties of a rare-earth compound have never been reproduced with one single method. For instance, none of the methods listed above reproduce the valence band spectrum of the materials they have been applied to, and instead of reproducing measured multiplet features they all result in more or less narrow energy bands. In addition, most of these methods have had difficulties in reproducing magnetic properties that are expected from a Russel-Saunders ground-state. The development of the Hubbard I approximation (HIA)~\cite{HIA} to treat the $4f$ shell of rare-earths is the focus of this investigation, and although it has been used in the past to calculate spectroscopic properties~\cite{Lebegue2006_light,Lebegue2006_heavy}, we examine here its capability to describe all salient aspects of rare-earth materials. Hence our study involves the cohesive properties, as evidenced by the equilibrium volume, bulk modulus and structural stabilities, as well as the magnetic properties, as revealed by the spin moments, orbital moments, and interatomic exchange parameters. The latter have also been compared with results obtained by treating the $4f$ electrons as core-like, which were in turn used to calculate ordering temperatures, magnon spectra and to investigate the stability of the ferromagnetic state. Finally, we used the Hubbard I approximation to address the spectroscopic data of the valence band (both occupied and unoccupied states). We reproduce with good accuracy a wide range of reported properties of the most important testing-ground of rare-earth materials, namely the elements. Based on our results we  argue that the HIA can be the basis of future investigations  addressing rare-earth based materials.



\section{Method}

The $[spd]$ electrons in the lanthanides are truly itinerant electrons and are well described by Kohn-Sham density functional theory (KS-DFT) in either the local density approximation (LDA) or the generalized gradient approximation (GGA). On the contrary, the $4f$ electrons are very localized and cannot be described by approximated exchange correlation functionals. In this article we use two methods to tackle this problem.  We mainly use the Hubbard I approximation, which combines the atomic multiplets of the $4f$ electrons with an LDA/GGA description of the itinerant conduction electrons. This method is explained in Sec.~\ref{sec:HIA} and will be used for an improved description of the cohesive properties, for the structural stabilities, for the ground-state magnetic properties and for the spectral properties. The exchange parameters $J_{ij}$ are calculated by means of the magnetic force theorem~\cite{lichtenstein-exch,PhysRevB.61.8906}, which can be easily combined with the Hubbard I approximation. However, this approach also involves a few technical ambiguities and eventually additional parameters (see below). Therefore, we preferred to address the exchange parameters by treating the $4f$ electrons as core states within the Kohn-Sham formalism (see Sec.~\ref{sec:SML}). Naturally we compared this approach to the HIA. Notice that the underlying assumption behind the $4f$-as-core approach and the Hubbard I approximation is the same, i.e. the lack of hybridization between the $4f$ states at a given site, and the rest of the valence electrons, including $4f$ states at other sites. In a sense, treating the $4f$ electrons as core states can be seen as poor man's version of the Hubbard I approximation. The exchange parameters were then used to calculate measurable quantities as the magnon spectra and the ordering temperatures ($T_{N/C}$'s) using atomistic spin dynamics (ASD) (Sec.~\ref{sec:SD}) and Monte Carlo (MC) simulations combined with the cumulant crossing method (CCM). An overview of the physical quantities we are going to investigate, as well as the method used to obtain them, are reported in Table~\ref{tab:methods}.

\begin{table}[htdp]
\caption{Investigated properties and employed methods. \label{tab:methods}}
\begin{center}
\begin{tabular}{lll}
\hline
\hline
Properties 	&Quantities 	& Method\\
\hline
Cohesive		&Volume, bulk modulus 			&HIA \\
                        &Structural stability				&HIA\\
Magnetic		&Magnetic moments			&HIA\\
				&Exchange parameters			& $f$ in core, HIA\\
				& \hspace{0.5cm}\rotatebox[origin=c]{180}{$\Lsh$} Ordering temperature&\hspace{0.3cm}\rotatebox[origin=c]{180}{$\Lsh$} MFA, MC\\
				&  \hspace{0.5cm}\rotatebox[origin=c]{180}{$\Lsh$} Fourier transforms	    &\hspace{0.3cm}\rotatebox[origin=c]{180}{$\Lsh$} ASD\\
				& \hspace{0.5cm}\rotatebox[origin=c]{180}{$\Lsh$} Magnon spectra& \hspace{0.3cm}\rotatebox[origin=c]{180}{$\Lsh$} ASD\\
Spectral 		& Photo-emission spectra  & HIA	\\
\hline	
\hline			
\end{tabular}
\end{center}
\end{table}%
 
\subsection{Hubbard I Approximation \label{sec:HIA}}

The main idea of the Hubbard I approximation is to combine the many-body structure of the $4f$ states, given by the atomic multiplets, with the broad Kohn-Sham bands of the delocalized valence electrons. This is done in the typical framework of the combination of DFT with the dynamical mean-field theory (DMFT), i.e. by using the DFT+DMFT Hamiltonian~\cite{HIA,kotliar06rmp78:865}. The latter is obtained in two steps. First the DFT Hamiltonian is projected onto the $4f$ orbitals (or more in general on those orbitals not correctly described in LDA/GGA). This projection results in a local Hamiltonian $\hat{H}_{LDA}^{\text{loc}}$. Second, a Hubbard interaction term $U$ is added explicitly to $\hat{H}_{LDA}^{\text{loc}}$ to describe the strong on-site Coulomb interaction between the $4f$ electrons.
The combined local Hamiltonian reads
\begin{multline}\label{eq:LDA+U}
\hat{H}=\hat{H}_{LDA}^{\text{loc}}  -\hat{H}_{DC}+\\
 +\frac{1}{2}\sum_{\mathbf{R}}{\sum_{\xi_1, \xi_2, \xi_3, \xi_4}
 {U^{\phantom{\dagger}}_{\xi_1, \xi_2, \xi_3, \xi_4} 
 \hat{c}^{\dagger}_{\mathbf{R},\xi_1}   
  \hat{c}^{\dagger}_{\mathbf{R},\xi_2} 
   \hat{c}^{\phantom{\dagger}}_{\mathbf{R},\xi_4} 
    \hat{c}^{\phantom{\dagger}}_{\mathbf{R},\xi_3} 
 }} 
\end{multline}
where $\mathbf{R}$ are Bravais lattice site vectors and $\hat{H}_{DC}$ is the so-called double counting (DC) term. This term should remove from $\hat{H}_{LDA}^{\text{loc}}$ those contributions that are due to $U$ and are incorrectly described in LDA/GGA. The orbital index $\xi$ labels the subset of  the ``correlated orbitals'' and in our case corresponds to atomic quantum numbers $n=4$, $l=3$, $m$ and $\sigma$.

The effective Hubbard model defined by Eq.~\eqref{eq:LDA+U} can be solved by means of DMFT~\cite{kotliar06rmp78:865}, i.e. through a mapping onto a single impurity Anderson model, under the condition of conserving the local Green's function. The bare Green's function of the correlated orbitals $\xi$ for the effective impurity model is 
\begin{equation}\label{eq:unpG}
\mathcal{\hat{G}}^{0}_{\text{imp}}(i\omega_n) 
= \frac{1}{(i\omega_n+\mu){{\hat{1}}}-\hat{H}_{LDA}^{\text{loc}}-\hat{\Delta}(i\omega_n)}
\end{equation}
where we have also introduced the fermionic Matsubara frequencies $\omega_n$. For sake of simplicity, $\mu$ contains the chemical potential and the double counting correction. $\Delta(i\omega_n)$ is instead the hybridization function between the $4f$ orbitals and the local environment around them. The Dyson equation gives the local Green's function in the impurity model
\begin{equation}
\hat{G}_{\text{imp}}(i\omega_n)=
\frac{1}{\mathcal{\hat{G}}^{0}_{\text{imp}}(i\omega_n)^{-1}-\hat{\Sigma}_{\text{imp}}(i\omega_n)}
\end{equation}
where $\hat{\Sigma}_{\text{imp}}(i\omega_n)$ is the self-energy function. To complete the mapping procedure, the hybridization function $\hat{\Delta}(i\omega_n)$ for a resulting self-energy $\hat{\Sigma}(i\omega_n)$ should be chosen such that $\hat{G}_{\text{imp}}(i\omega_n)$ reproduces the local Green's function at a single site in the Hubbard model $\hat{G}_{\mathbf{R}\mathbf{R}}(i\omega_n)$. In DMFT, one approximates the lattice self-energy as:
\begin{equation}\label{eq:sigma}
\hat{\Sigma}_{\mathbf{R}\mathbf{R}'}(i\omega_m)=
\delta_{\mathbf{R}\mathbf{R}'}\hat{\Sigma}_{\text{imp}}(i\omega_m)
\end{equation}
The self-energy is therefore local, or in other words $k$-independent.

The effective impurity model arising in DMFT can be solved by various techniques~\cite{kotliar06rmp78:865}. In HIA a solution of the effective impurity model is obtained by neglecting the coupling between the bath and the impurity, i.e. the hybridization in Eq. \ref{eq:unpG}. This boils down to approximating the self-energy in the impurity problem (Eq.~\eqref{eq:sigma}) by the atomic self-energy $\hat{\Sigma}_{\text{at}}(i\omega_m)$, which still retains the proper multi-configurational description. For the lanthanides this is a meaningful approximation, since the $4f$ orbitals are chemically almost inert and are very much atomic-like. As we mentioned above, this approach is reminiscent of treating the $4f$ electrons as core-like states, as both approaches are based on the assumption of negligible hybridization of the $4f$ electrons. However, in HIA the $4f$ states are real valence electrons, possessing a proper multiplet spectrum, and can therefore be used in a much more flexible way. Although the self-energy is atomic-like, the feedback effect generated when the self-energy is taken back into the lattice through Eq.~\eqref{eq:sigma} leads to some (small) hybridization effects. Once the self-energy has been determined, the computational cycle can be closed by adjusting the chemical potential and determining the full lattice Green's function $\hat{G}_{\mathbf{R}\mathbf{R'}}(i\omega_n)$. This can be used to update the electron density and create a new $\hat{H}_{LDA}$ in Eq.~\eqref{eq:LDA+U}. This is the so-called charge self-consistent cycle. When both the electron density and the local self-energy have been converged, the most important physical properties can be obtained through the lattice Green's function. The spectral function is calculated as
\begin{equation}
\rho(\omega)= -\frac{1}{\pi}\text{Im} \left[ G(\omega+i\delta)\right] \quad \text{for }\delta \rightarrow +0
\end{equation}

In this study we use the DFT+DMFT approach and the HIA as implemented in the Full-Potential Linear Muffin-Tin Orbital (FP-LMTO) code \texttt{RSPt}~\cite{RSPtbook,Patrik1,granas12cms55:295}. The construction of the correlated orbitals is based on the muffin-tin heads, and is described extensively in Ref.~\cite{grechnev07prb76:035107}. The total energies have been evaluated by following the approach of Ref.~\cite{dimarco09prb79:115111}.

\subsection{DFT with $4f$ as core electrons for exchange parameters \label{sec:SML}}
Most of the results concerning the exchange parameters $J_{ij}$ have been obtained by using the expression for the exchange parameters of the Heisenberg model derived in Ref.~\cite{lichtenstein-exch}. In these simulations the $4f$ electrons were treated as core ones. 
This is justified by the fact that these states are to a large extent chemically inert due to their very localized nature, and the Fermi surface of the two treatments is expected to be similar. 
In this way the orbitals are still not described correctly by the approximate exchange correlation functionals, but at least they do not hybridize with the orbitals, centered at the neighbouring atoms. 
Therefore they do not influence the bonding nor spuriously contribute with the direct exchange, which appears due to the overestimation of the wavefunctions extension (see e.g. Ref.~\cite{OE-Gd-95}).
There are a few technical reasons why this approach is preferable to the direct use of the expression for the exchange parameters derived in Ref.~\cite{lichtenstein-exch}, in the HIA. These issues will be discussed more in detail in Section~\ref{sec:compdet_exch}.

\subsection{Atomistic Spin dynamics and Monte Carlo\label{sec:SD}}
The calculated exchange parameters were used as an input for the ASD simulations.
The formalism is described in Ref.~\cite{Ref-ASD1} and is implemented in the \texttt{UppASD} package~\cite{Ref-ASD2}.
Within this approach the motion of the magnetic moments is described semi-classically.
The spins are treated as classical vectors, exposed to different types of magnetic interactions, which have quantum origin.
In our study we have employed the following spin Hamiltonian:
\begin{equation}\label{eq:SpinH}
H = - \sum_{ i \neq j } J_{ij} \; \mathbf{e}_i \cdot \mathbf{e}_j \; + \; K \sum_{i} (\mathbf{e}_{i} \cdot \mathbf{e}_{K})^2 \;,
\end{equation}
where $(i,j)$ are atomic indices, $\mathbf{e}_{i}$ is the unit vector along the direction of the atomic spin moment at the site $i$, $J_{ij}$ is Heisenberg exchange coupling parameter and $K$ is the strength of the anisotropy field pointing along the direction of $\mathbf{e}_K$. 
The dynamics of the atomic spins at finite temperature is governed by Langevin dynamics and is described in the Landau-Lifshitz-Gilbert (LLG) equation, which reads:
\begin{equation}
\begin{split}
\frac{\partial \mathbf{s}_i}{\partial t} = - \frac{\gamma}{1+\alpha^2_i} \; \mathbf{s}_i \; \times \; [\mathbf{B}_i + \mathbf{b}_i(t)] \; + \\ 
& \hspace{-4cm} - \frac{\gamma\; \alpha_i}{s_i (1+\alpha^2_i)} \mathbf{s}_i \; \times \; \{ \mathbf{s}_i \; \times \; [\mathbf{B}_i + \mathbf{b}_i(t)] \} \;,
\end{split}
\label{LLG}
\end{equation}
where $\mathbf{s}_i = s_i \mathbf{e}_i$ is the spin moment vector at the site $i$, $\gamma$ is the gyromagnetic ratio and $\alpha_i$ denotes a dimensionless site-dependent damping parameter, which gives rise to the energy dissipation and eventually brings the system to a thermal equilibrium. 
The effective field that the spins are exposed to, consists of an intrinsic field ($\mathbf{B}_i = - \partial H / \partial \mathbf{s}_i$) and a stochastic magnetic Gaussian-shaped field $\mathbf{b}_i$. 
The latter one is introduced to mimic the effects of thermal fluctuations and is directly related to $\alpha_i$ parameter. 
Once Eq.~\eqref{LLG} is solved, the time evolution of every spin moment ($\mathbf{s}_i(t)$) is known. This allows to evaluate any dynamical property of the system, such as spin wave excitation spectrum.
More information about the dispersion relation calculation can be found in Refs.~\cite{Ref-ASD2, Ref-ASD3}.



\section{Computational details\label{sec:comp_det}}
\subsection{Cohesive properties, structural stabilities and ground-state magnetic moments and photo-emission spectra\label{sec:compdet_coh}}
The cohesive properties, the structural stabilities, the ground-state magnetic moments, and the spectral functions were calculated using the Hubbard I approximation as implemented in \texttt{RSPt}~\cite{RSPtbook}. The FP-LMTO basis was constructed as described in the following. The $5s$ and $5p$ states were included as semi-core states with two spherical Hankel functions at $0.3$~Ry and $-2.3$~Ry. The linearization energy, at which the radial Koelling-Harmon scalar relativistic equation was solved,  was set to the center of the band. The completeness of the basis was found to significantly increase by including also the $5f$ basis functions (with the same parameters as $5s$ and $5p$). The valence electrons were described with $6s$, $6p$, $5d$ and $4f$ basis functions. The $6s$ and $6p$ states were described with three spherical Hankel functions at  $0.3$~Ry,  $-2.3$~Ry and $-1.5$~Ry and the linearization energy was chosen such that the functions are orthonormalized to the $5s$ and $5p$ basis functions. The $5d$ and $4f$ states were described with two spherical Hankel functions at $0.3$~Ry and $-2.3$~Ry. The linearization energy of the $4f$ was set to the center of the band. For the divalent elements, it was found to be important to set the linearization energy of the $5d$ also to the centre of the band. The spin-orbit coupling (SOC) was taken into account in a perturbative way~\cite{RSPtbook}. Spin-polarization was only taken into account for the magnetic properties, as in fact HIA is capable to describe the proper paramagnetic phase of a material~\cite{kotliar06rmp78:865}. The exchange correlation functional was treated in both LDA~\cite{PW92} and GGA~\cite{AM05,PW92}.

The ground-state properties were determined by assuming the experimental lattice structure, except for the dhcp structures which were approximated by fcc. Thus, Ba was calculated in bcc structure, Ce to Pm in the fcc structure, Sm in the Sm-structure (9R), Eu in bcc structure, Gd to Tm and Lu in the hcp structure and Yb in the fcc structure. For hcp structures we assumed the experimental $c/a$ ratio. Especially for the bulk modulus it was found to be important to use a large number of $k$-points. The reciprocal space was sampled with about $10000$ $k$-points in the full Brillouin zone, for each structure.

The $U$ matrix in Eq.~\eqref{eq:LDA+U} was constructed from the Slater parameters $F^0$, $F^2$, $F^4$ and $F^6$~\cite{kotliar06rmp78:865}. 
$F^0$, which corresponds to the Hubbard $U$, was treated as a parameter and fixed at a value of $U=7$~eV throughout the whole series. This choice is based on calculated values found in literature~\cite{Nilsson_U,Pourovskii_U,Marel_U} as well as on the agreement between previous calculations and experimental spectra~\cite{Lebegue2006_light,Lebegue2006_heavy}. Other choices of $U$ are of course also possible, such as calculated values obtained with cLDA, cRPA or linear response methods. This involves however some additional technicalities, such as renormalizing the Kohn-Sham Hamitonian~\cite{PhysRevLett.109.126408} and having the same basis for the correlated orbitals. In Ref.~\cite{Kolorenc} is has been argued that the Hubbard $U$ in the HIA should be decreased compared to the $U$ in full DMFT for the light actinide dioxides to mimic the hybridization. However, for the lanthanides, the hybridization is at least 1 order of magnitude smaller than for the light actinide dioxides, so we predict the discrepancy to be negligible.
The other Slater parameters were directly calculated at experimental volume and then screened by an empirical factor~\cite{Patrik1,Patrik}: $F^2\rightarrow0.92F^2$ and $F^4\rightarrow0.97F^4$. The correction over the values obtained directly from the KS wave functions is tiny, but this small change is noticeable in the spectroscopic data. The parameters were found to depend only slightly (meV) on the exchange correlation functional (LDA/GGA) or on the various double-counting schemes. From the Slater parameters, we can evaluate the Hund's $J$, whose values are reported in Table \ref{tab:J}. A slightly increasing $J$ is found across the series, starting from $J \approx 0.5$~eV for Ba to $J \approx 1.3$~eV for Lu, as can be expected from Slater rule's ~\cite{SlaterRules}. This increase reflects the contraction of the radial part of the $f$ wave-functions due to the increasing core charge, which results in an increasing importance of the electron-electron repulsion. We tested the influence of changing $U$ or $J$ on the volume and bulk modulus for Nd and Tm. The effect on these materials is rather small; a change of about 30\% in one of the parameters results in a difference in volume of about $0.3~$\AA$^3$ and a difference in bulk modulus of about $1$GPa.

\begin{table}[htdp]
\caption{Calculated Hund's $J$'s for the lanthanides.\label{tab:J}}
\begin{center}
\begin{tabular}{lll  lll}
\hline
\hline
$Z$ \hspace{0.3cm} &Name	&$J$~(eV)\hspace{1cm} &$Z$ \hspace{0.3cm} &Name	&$J$~(eV)\\
\hline	
57	&La		&0.75				&65	&Tb		&1.08	\\
58	&Ce		&0.79				&66	&Dy		&1.11\\
59	&Pr			&0.84				&67	&Ho		&1.15	\\
60	&Nd		&0.89				&68	&Er		& 1.23	\\
61	&Pm		&0.93				&69	&Tm		&1.21	\\
62	&Sm		&0.97				&70	&Yb		&1.16	\\
63	&Eu		&0.91				&71	&Lu		&1.27	\\	
64	&Gd		&1.05				&		&			&\\
\hline
\hline
\end{tabular}
\end{center}
\end{table}%

The double counting correction $\hat{H}_{DC}$ in Eq.~\eqref{eq:LDA+U} was chosen as in the fully localized limit, but replacing the occupation $N$ of the $4f$ shell with the closest integer number $N_{at}$, in the spirit of Ref.~\cite{Pourovskii_U}. This results in a total energy correction:
\begin{equation}\label{eq:DCFLL}
E_{DC}=\frac{1}{2}UN_{at}\left(N_{at}-1\right)-J\frac{N_{at}}{2}\left(\frac{N_{at}}{2}-1\right)
\end{equation}
This choice of double counting, which is unique to HIA, should be accompanied by evaluating the Galitskii-Migdal contribution to the total energies~\cite{dimarco09prb79:115111} directly into the atomic problem. 
Due to that in our implementation of HIA, this contribution is instead evaluated in the global LDA basis~\cite{granas12cms55:295}, the total energy correction in Eq.~\eqref{eq:DCFLL} should be renormalized to take into account the contribution of the hybridization. At first order, this can be simply done by replacing the factors $N_{at} N_{at}$ in Eq.~\eqref{eq:DCFLL} with $N N_{at}$. A more elaborate explanation of this correction can be found in Ref.~\cite{licInka}.

For the magnetic properties, the double counting issue is much more complicated, as recently outlined in Ref.~\cite{peters14prb89_205109}.
When combining spin-polarized DFT with HIA (or DMFT in general), $\hat{H}_{DC}$ should correct for the $f$-$f$ exchange already taken into account in the DFT part and again in $U$. However, it should not correct for the $f$-$d$ exchange, since this is not taken into account for by the Hubbard $U$. The problem is that one can not disentangle the two, and therefore any choice becomes an unjustified guess. To avoid this problem, we prefer to obtain the magnetic properties by applying HIA to non spin-polarized DFT, after a small field has been applied to break the spin symmetry. This results in a local Hamiltonian possessing a tiny polarization, which is enough to lead to a fully spin-polarized self-energy. Given that the $\hat{H}_{LDA}$ contains only the (vanishing) spin polarization due to the tiny external field, the double counting correction has to be non spin-polarized. This is obtained by taking the average of $\hat{H}_{DC}$ over both spin channels. The correction to average can be again constructed through the fully localized limit~\cite{kotliar06rmp78:865}.

For calculating the spectra the double counting correction was chosen such that the number of $f$ electrons is correct and that the position of the first occupied (or unoccupied) peak is aligned to the corresponding experimental peak in the direct (or inverse) photoemission spectrum. This choice makes it easy and insightful to compare the theoretical and experimental spectra. We evidently compared this to the spectra calculated with the fully localized limit double counting. We mainly see a rigid shift of the $4f$ multiplet with respect to the $[spd]$ density of states, as is shown in Appendix.~\ref{app:LDA_GGA_FP_FLLL}. Notice that this is in good agreement with a recent work on the double counting corrections for the DFT+DMFT method~\cite{haule_arxiv}.

\subsection{Exchange parameters\label{sec:compdet_exch}}
We have emphasized above that it is rather difficult to obtain a proper spin-polarized double counting correction in DFT+DMFT. For most materials, and for most properties, this problem can be circumvented by starting the DMFT cycle from non spin-polarized DFT and let the whole magnetism arise from the $U$ term in Eq.~\eqref{eq:LDA+U}. Obviously this procedure cannot be performed to reach full charge self-consistency, due to that the KS Hamiltonian (without the tiny magnetic field) is only going to be non spin-polarized at the first iteration. Unfortunately, for the heavy REs, the effects of charge self-consistency are very important, due to that native LDA or GGA calculations tend to overestimate the occupation of the $4f$ shell of as much as 0.5 electrons. Therefore, this procedure is not applicable to the calculation of the interatomic exchange parameters, which are very sensitive to this redistribution of electron density. A charge self-consistent calculation within the HIA can be done by fine-tuning the parameters involved in the simulation, as illustrated in Appendix~\ref{app:cscHIA}. This Appendix contains a comparison between results obtained with HIA in a single shot calculation and in a charge self-consistent calculation. The results emphasize the importance of a charge self-consistent description of the density.
In the end of Sec.~\ref{res:mag} a comparison between magnon spectra obtained with HIA exchange parameters and with $4f$-as-core exchange parameters is made. We show there that the overall shapes of the magnon spectra  are similar.
Since we did not find any large qualitative differences between these two types of calculations, we decided to simulate the properties related to the magnetic excitations using the $4f$-as-core setup, which is computationally more expedient, as mentioned above.
Thus, for this study we have treated the $4f$ states as non-hybridizing atomic states, allowing for their spin-polarization. 
In this way the $4f$ electrons produced an effective magnetic field within the muffin-tin (MT), influencing the itinerant $[spd]$ levels.
Throughout the whole paper, whether the $4f$ electrons were present in the valence or not, the exchange parameters were always computed between the available valence states, projected on the head of the corresponding MT sphere.
Every state was described with the set of three basis functions having different kinetic energy tails. The tail energies used where the same as those used in Sec.~\ref{sec:compdet_coh} for the cohesive properties, the ground state magnetic properties and the spectra. 
For each RE element, we have chosen the MT radii such that the spheres were almost touching.
We have adopted the experimental lattice constant for each metal.
The states at the Fermi level were smeared using the Fermi-Dirac scheme with an effective temperature of 2.5 mRy ($\approx$400 K).

It is worth mentioning that we had to utilize a very dense $k$-point grid (38$\times$38$\times$24 divisions) to correctly describe the interaction with the far-distant neighbors (which were found to be important). Further details of the implementation of the magnetic force theorem in \texttt{RSPt} code can be found in Ref.~\cite{jijs-rspt}. We finally note that these calculations were performed without including SOC. The $5d$-band width, that is mainly responsible for mediating the exchange between the localized $4f$-bands, is much larger than the spin-orbit coupling strength.

\subsection{Ordering temperatures and magnon spectra}

Using the values of the exchange parameters and magnetic moments from DFT, we estimated the ordering temperature $T_{N/C}$ for the heavy RE metals by means of Monte Carlo simulations and using the cumulant crossing method~\cite{RefCum}. The Hamiltonian used to estimate the ordering temperature is described by Eq.~\ref{eq:SpinH} but neglecting the anisotropy term.
The results were compared with the simple estimates, based on mean-field approximation (MFA).
We underline the fact that due to the long-range nature of the magnetic couplings, a relatively large number of exchange interactions was required in order to sufficiently converge the $T_{N/C}$ value. 
For most of the simulations we had to include the $J_{ij}$'s with all nearest neighbors (NN) within the distance of 5.57 lattice parameters, which corresponds to taking into account the nearest 1098 neighbors of each atom.
The magnon spectrum of Gd was simulated with ASD, using a low-temperature experimental value of the uniaxial anisotropy constant $K_1 = 2.5$~$\mu\text{Ry}$ (Ref.~\cite{RefGd-aniso}).
A simulation box containing 50$\times$50$\times$50 sites with periodic boundary conditions was adopted. The temperature was set to T=78 K, $\alpha = 0.001 $ and the exchange interactions with all neighbours within $d_{max}\leq5.57a$, where $a$ is the lattice parameter, were taken into account.



\section{Results}
We will present first the cohesive properties \textit{i.e.} the equilibrium lattice parameter and the bulk modulus in Sec. \ref{sec:cohesive properties}. Second we discuss the structural stabilities in Sec.~\ref{sec:strucstab}. Third the magnetic properties \textit{i.e.} the magnetic ground-state, the exchange parameters, the ordering temperatures, the Fourier transform of the exchange parameters and the magnon spectra are presented in Sec. \ref{sec:magnetic properties}. Fourth the calculated valence band (inverse) photoemission spectra are discussed in Sec. \ref{sec:spectra}.

\subsection{Cohesive properties\label{sec:cohesive properties}}
To obtain the cohesive properties, the energy was calculated for several lattice parameters with the Hubbard I approximation in a charge self-consistent mode. The energy versus lattice parameter was calculated from -3\% to +3\% around the theoretical equilibrium lattice parameters for the structures described in Sec. \ref{sec:comp_det}. These curves were fitted with the Birch-Murnaghan equation of state ~\cite{Birch}.

\begin{figure}
     \centering
         \includegraphics[width=\columnwidth]{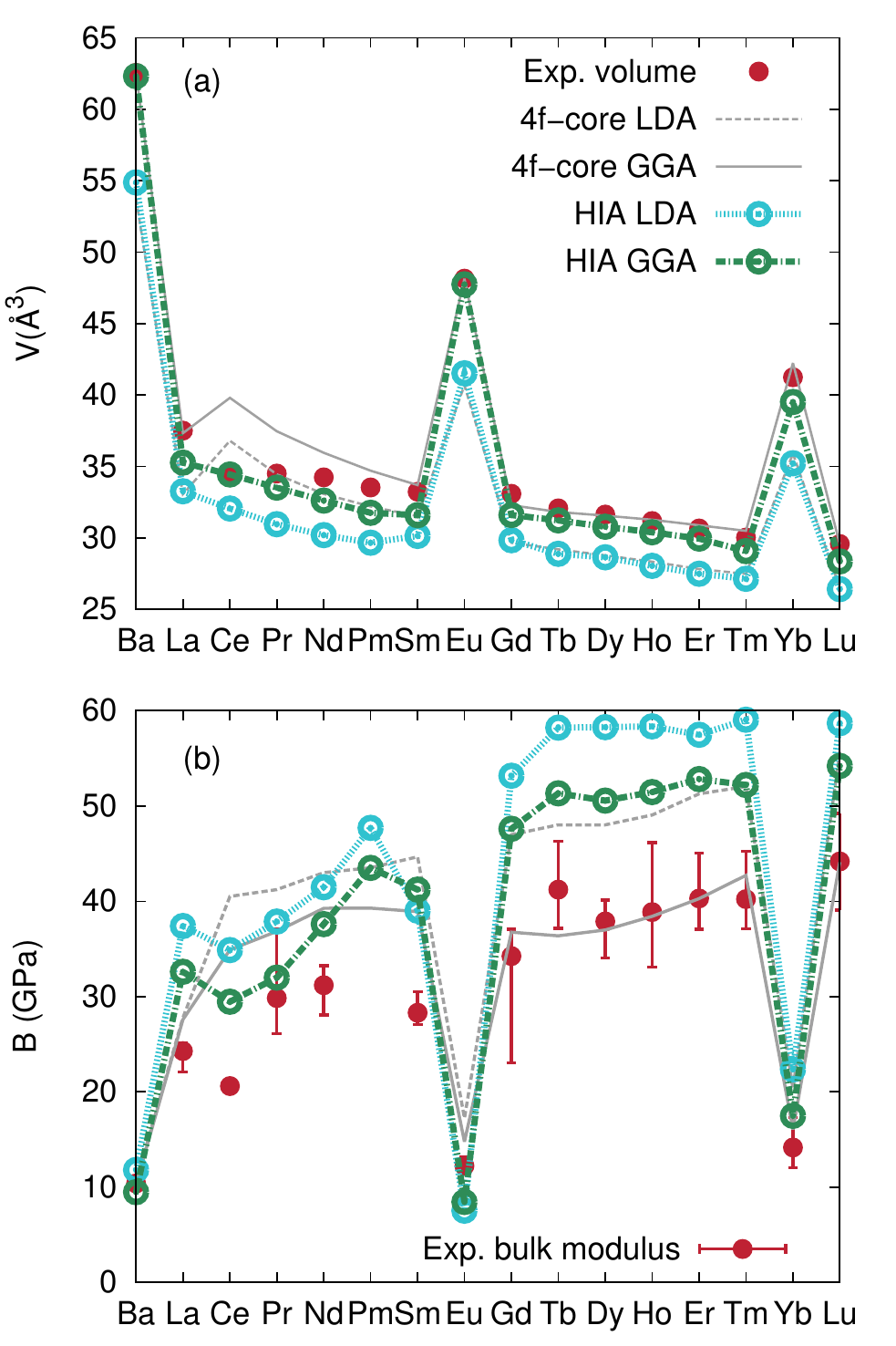} 
  \caption{Cohesive properties: (a) Equilibrium atomic volume and (b) bulk modulus for all elements of the REs series. The red solid dots are experimental data from Ref.~\cite{exp_V} for the volumes and from Ref.~\cite{exp_B} for the bulk moduli. The light-grey lines are volumes calculated with the $4f$'s treated as core electrons in LDA and GGA by A. Delin {\it{et al.}} in Ref.~\cite{Delin1998}. The open dots are the present calculations with the Hubbard I approximation in LDA and GGA (see labels in the plots). 
\label{fig:V_B}}
\end{figure}

In Figure~\ref{fig:V_B} we present our results on the equilibrium volume and the bulk modulus for the entire series of the lanthanides. Figure~\ref{fig:V_B}(a) shows the calculated equilibrium volumes per atom compared to experimentally measured volumes. The trend across the series is quite well captured, both with the LDA and the GGA functionals. However the LDA functional displays the usual underestimation of the volume, which is caused by an over binding of this functional of itinerant $[spd]$-electrons. The lanthanide contraction is slightly overestimated by both functionals. This overestimation of the contraction may be dependent on the value of $U$. For the light REs, a more pronounced overestimation of the lanthanide contraction had previously been found by A. Delin et al. in Ref.~\cite{Delin1998}. In that study the $4f$ electrons were treated as core electrons. P. S\"oderlind et al. treated the $4f$ electrons in the REs band-like but with spin and orbital polarization~\cite{soderlind} and found a quite pronounced increase in the volume across the light rare earths, which is a trend opposite to experiment. P. Strange et al.~\cite{PStrange}, who treated the REs using SIC, found a slight increase in the volume across the light rare earths. Thus the trend in equilibrium volume for the early elements is captured better by the Hubbard I approximation than by treating the $4f$ electrons as core electrons, using the orbital polarization method or the SIC method. 
For the late lanthanides the HIA captures the trend very well, but even the GGA functional underestimates the equilibrium volumes slightly. In Ref.~\cite{Delin1998} a small underestimation of the contraction was found.  In Refs.~\cite{soderlind,PStrange} quite good agreement was found for the volume of the heavy REs, although some small unexpected increases are noticed (see also Appendix~\ref{app:compdiffmeth}). 

Figure~\ref{fig:V_B}(b) shows the bulk moduli (at the theoretical equilibrium volume) for the two functionals and the experimental values. The experimental values found in the literature for the bulk modulus differ quite a lot between different studies. We used values from Ref.~\cite{exp_B}, where the literature values (up to 1991) are averaged with error bars indicating the lowest and highest reported values.
Both functionals capture the trend in the bulk modulus quite well.
The relative difference between the bulk modulus for the divalent and trivalent elements is found correctly. Also the trend between the other consecutive elements is consistent with the experimental measurements. In all cases (except for Eu), the bulk modulus is overestimated. This can partially be attributed to the the slight underestimation of the volume. In Appendix~\ref{app:compdiffmeth} the HIA is compared to other methods.
The LDA calculations for Sm, Tb and Tm showed some problems in convergence due to the high number of $k$-points needed for an accurate calculation of the bulk moduli. However we increased the $k$-mesh for all elements until the results were reasonably stable.

\subsection{Structural Stability\label{sec:strucstab}}		
In the previous section we assumed the experimental crystal structure for most calculations. However it is interesting whether HIA actually can predict the crystal structure with the lowest energy. Although a full analysis of all relevant crystal structures of the lanthanides (fcc, hcp, dhcp and 9R) is outside the scope of this investigation, we give here an example of how well HIA reproduces structural properties of Ce, Pr and Nd. For this limited example, we compare the stability only of fcc, hcp and dhcp. We find in all three cases that the dhcp structure is about 10~meV lower in energy than the fcc structure, which is in turn 20~meV lower than the hcp structure. For Pr and Nd, the experimental crystal structure at ambient pressure and room temperature is the dhcp structure~\cite{jensmac,PhysRevB.11.2836}, in agreement with our calculations. With increasing pressure or temperature a transition to the fcc structure would occur~\cite{PhysRevB.11.2836}. For Ce, our predicted dhcp structure might at first sight seem surprising, since the experimentally predicted structure is fcc~\cite{jensmac}. Although our prediction is indeed wrong, it is not completely unexpected. In the phase diagram of Ce, at low pressure, there is a low temperature $\alpha$-phase (fcc), a medium temperature $\beta$-phase (dhcp) and around room temperature there is a $\gamma$-phase (fcc). Compared to the $\gamma$-phase, the $4f$ electrons in the $\alpha$-phase hybridize significantly more with the surrounding. Therefore the HIA is not able to describe this phase. If we therefore remove the $\alpha$-phase from the phase diagram, the expected phase at low temperature would be the $\beta$-phase. Hence, among the phases with entirely localized $4f$ electrons in Ce, the HIA  reproduces the expected crystal structure.

\subsection{Magnetic properties \label{sec:magnetic properties}}
\subsubsection{Hund's rules, ground-state magnetic moments}
The Russell-Saunders coupling scheme is normally adopted to describe the $4f$ magnetism of the REs. The spins of the individual electrons are coupled to form a total spin $S$ by the exchange interaction and the individual orbital angular momenta are coupled to form a total orbital momentum $L$ by the Coulomb interaction. The state with the lowest energy is found from Hund's rules by maximizing $S$ and thereafter $L$. The total angular momentum $J$ is given by $J=|L\pm S|$, where the minus sign is used for less than half filled shells and the plus sign for more than half filled shells. 
The total magnetic moment due to the spin angular momentum is $\mu_S=2\mu_B\sqrt{S(S+1)}$, where $\mu_B$ is the Bohr magneton. The total moment due to the orbital angular momentum is $\mu_L=\mu_B\sqrt{L(L+1)}$. These moments precess around the direction of $J$, therefore the magnetic moment due to the total angular momentum looks slightly more complicated: $\mu_J=g_J\mu_B\sqrt{J(J+1)}$, where $g_J$ is the Land\'e factor~\cite{jensmac}. The saturation moment is given by the projection of $\boldsymbol{\mu}_J$ on the direction of the applied field, the $\hat{\mathbf{z}}$-axis and is given by $\mu_f=g_JJ$.
In our case, these $S$, $L$ and $J$ are the spin, orbit and total angular momentum originating from the $4f$ electrons. The conduction electrons ($[spd]$) are known to have a measurable contribution to the total moment, with a value up to $0.65$~$\mu_B$ for Gd that has the largest $[spd]$ contribution of the entire series~\cite{STICHT1985529}, as will be discussed below. In addition, the conduction states are very important for mediating the exchange interaction between the $f$-moments on different sites.
 
 \begin{figure}[!t]
     \centering
      \includegraphics[width=\columnwidth]{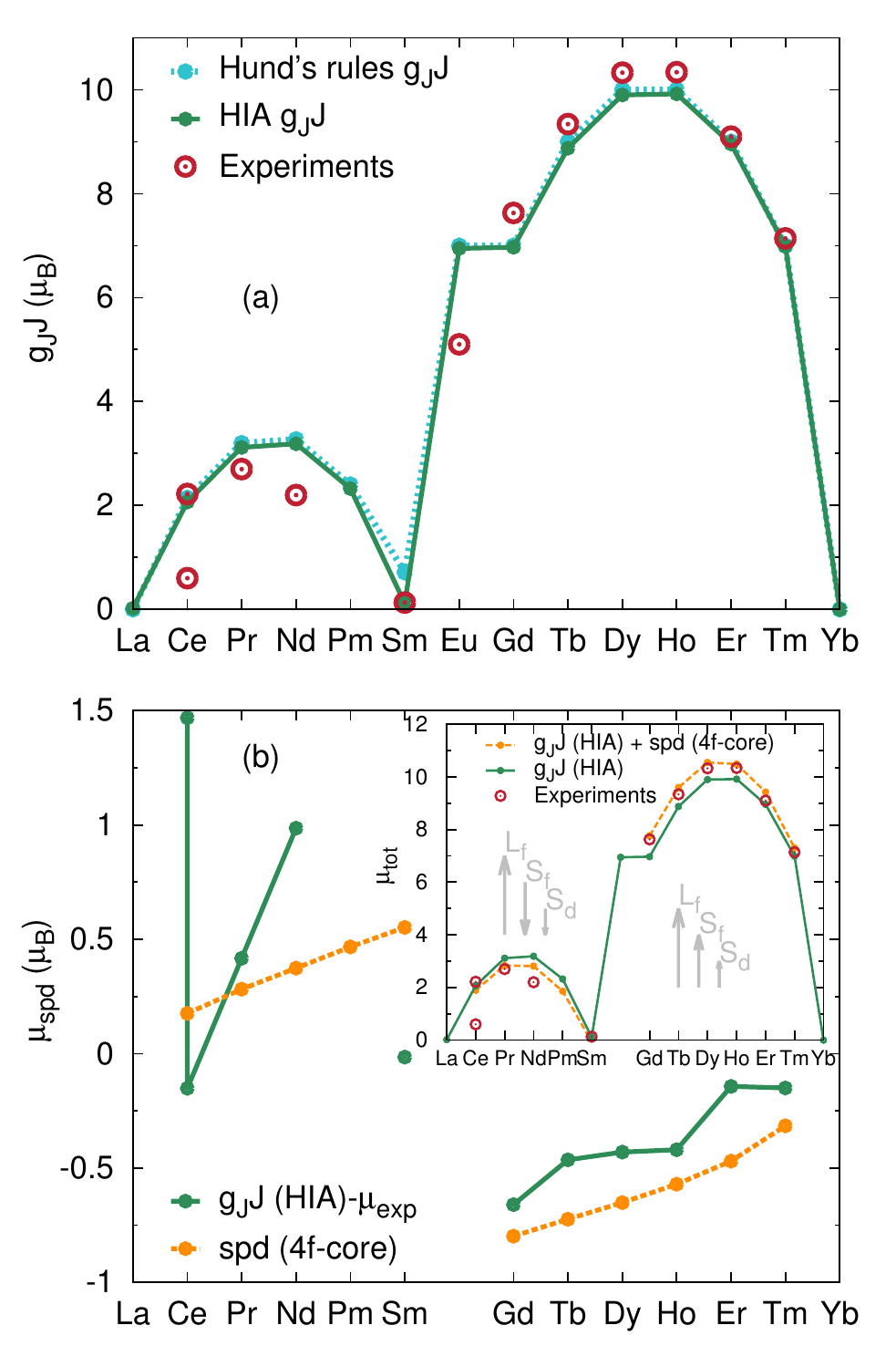} 
\caption{Magnetic ground-state properties. (a): the $4f$ saturation moment ($g_JJ$, where $g_J$ is the Land\'e factor) following from Hund's rules (blue line) and the $4f$-moment calculated with HIA (green line). The red dots are experimentally measured saturated moments taken from the Table 1.6 of  Ref.~\cite{jensmac} and Chapter 6 of Ref.~\cite{exp_mom}. The highest experimental value for Ce is taken from the more recent work of Ref.~\cite{drymiotis}. (b): the $4f$-moment calculated with HIA minus the experimental moment (green line) and the $[spd]$-moment calculated with the $4f$ in the core model (orange line). In the inset the $4f$-moment calculated with HIA is added to this $[spd]$-moment.\label{fig:M}}
\end{figure}
 
%

The magnetic moments were calculated at the experimental volume for the structures described in Sec.~\ref{sec:compdet_coh} in a ferromagnetic orientation. An external magnetic field of $0.05$~Ry/bohr was added to polarize the $f$ electrons, according to the arguments of Ref.~\cite{peters14prb89_205109}. The temperature was set at $T=0.1$~mRy in order to select only the ground-state and to avoid superpositions of different configurations. Tests with fields up to 50 times smaller in size, i.e. of about $0.001$~Ry/bohr were investigated, and resulted into the same $4f$-moments, provided that the temperature was also reduced accordingly. 
In Figure~\ref{fig:M}(a) we present the saturated magnetic moments, the projection of $\boldsymbol{\mu}_J$ on the $\hat{\mathbf{z}}$-axis, $\mu_f=g_JJ$. This saturation moment and the Land\'e factor are calculated both from the $S$, $L$ and $J$ as expected from Hund's rules (for Hund's rules $g_JJ$ ) and from the $S$, $L$ and $J$ that are obtained as expectation values of the corresponding angular momentum operators in the local Green's function (for HIA  $g_JJ$). The HIA approximation recovers the Hund's rule ground-state as the state with the lowest energy and the magnetic moments of the two different calculations are practically the same.  These similarities are partially expected since both Hund's rules and HIA correspond to an atomic-like picture, although HIA also includes crystal field effects. However, it is rewarding that a theory which does not rely on any assumption on couplings between spin and angular momenta, like HIA, results in the expected behavior. This is not self-evident, as is illustrated for example for TbN in Ref.~\cite{peters14prb89_205109}  where the problem of finding correct orbital moments of rare-earth systems using e.g. LSDA+U and LSDA was discussed.

One may note in Figure~\ref{fig:M}, a good agreement between the measured~\cite{jensmac,exp_mom} and the calculated moments for the whole series. The most striking differences are found for Ce, Nd and Eu. For Ce two very different experimental values for the saturation moment were found in literature. We expect that a small error in the calculated value  should arise from neglecting a still finite hybridization. In Ref.~\cite{exp_mom} is noticed that the measured magnetic moment for Nd is quite far from what is expected from Hund's rules. Moreover, the paramagnetic moment ($g_J\sqrt{J(J+1)}$) was measured to be $3.4\mu_B$, whereas the saturated moment ($g_JJ$) was measured to be $2.2 \mu_B$  for Nd in Ref.~\cite{jensmac}. These values do not correspond with one an other.  For Pm no experimental data are available, since it is radioactive. Ref.~\cite{exp_mom} reports that the experimental value of the magnetic moment for Eu does not correspond to the true saturation value. This is due to the fact that in Eu the moments form a spin spiral~\cite{exp_mom} and, if a field is applied, are lifted out of the plane to form a helix. However it is very hard to distort this helix and the magnetic field used in the experiment was not high enough to reach saturation. Apart from these cases, the agreement is good.\\
We notice, however, an overall overestimation of the calculated saturation moments compared to the experimental moments for the light rare earths and an overall underestimation for the heavy rare earths. This can be partially attributed to the $[spd]$ polarization. The $[spd]$ electrons contribute mainly with a spin moment that is parallel to the spin moment of the $4f$'s.  For the light rare-earth elements the magnetic moment is dominated by the orbital $f$ contribution, with a smaller spin $f$ contribution that is coupled anti-parallel to the orbital moment. This results in a reduced saturation moment when the $[spd]$ contribution is accounted for in the early rare-earth elements. For the heavy rare-earth elements, where the spin and orbital moment are parallel, including the $[spd]$ contribution results in a bigger saturation moment. The different contributions to the total magnetic moment are illustrated by the grey arrows in the inset of Figure~\ref{fig:M}(b). To investigate the contribution of a spin-polarized $[spd]$ band, we subtracted the experimental moment from the moment calculated with HIA. In Figure~\ref{fig:M}(b) we compared this difference with the $[spd]$ moment (as calculated when the $4f$ states were treated as core electrons). For the heavy lanthanides the $[spd]$-moment follows the same trend as the difference between the experimental moment and the total $4f$ moment from HIA. Hence the difference between theory and experiment in Figure~\ref{fig:M}(a) may be attributed primarily to the $[spd]$-derived magnetism. For the light lanthanides the situation is a little more complex, partially since there are several experimental values. In the inset of Figure~\ref{fig:M}(b) we added the $[spd]$ moment (calculated with the $4f$-in-the-core treatment) to the $4f$ moment calculated with HIA. For the late REs we find an excellent agreement, especially given the magnitude of the moments.  Europium was excluded from this comparison since the experimental value does not correspond to the true saturation value. 

\subsubsection{Exchange parameters $J_{ij}$}

\begin{figure}
     \centering
     \includegraphics[width=\columnwidth]{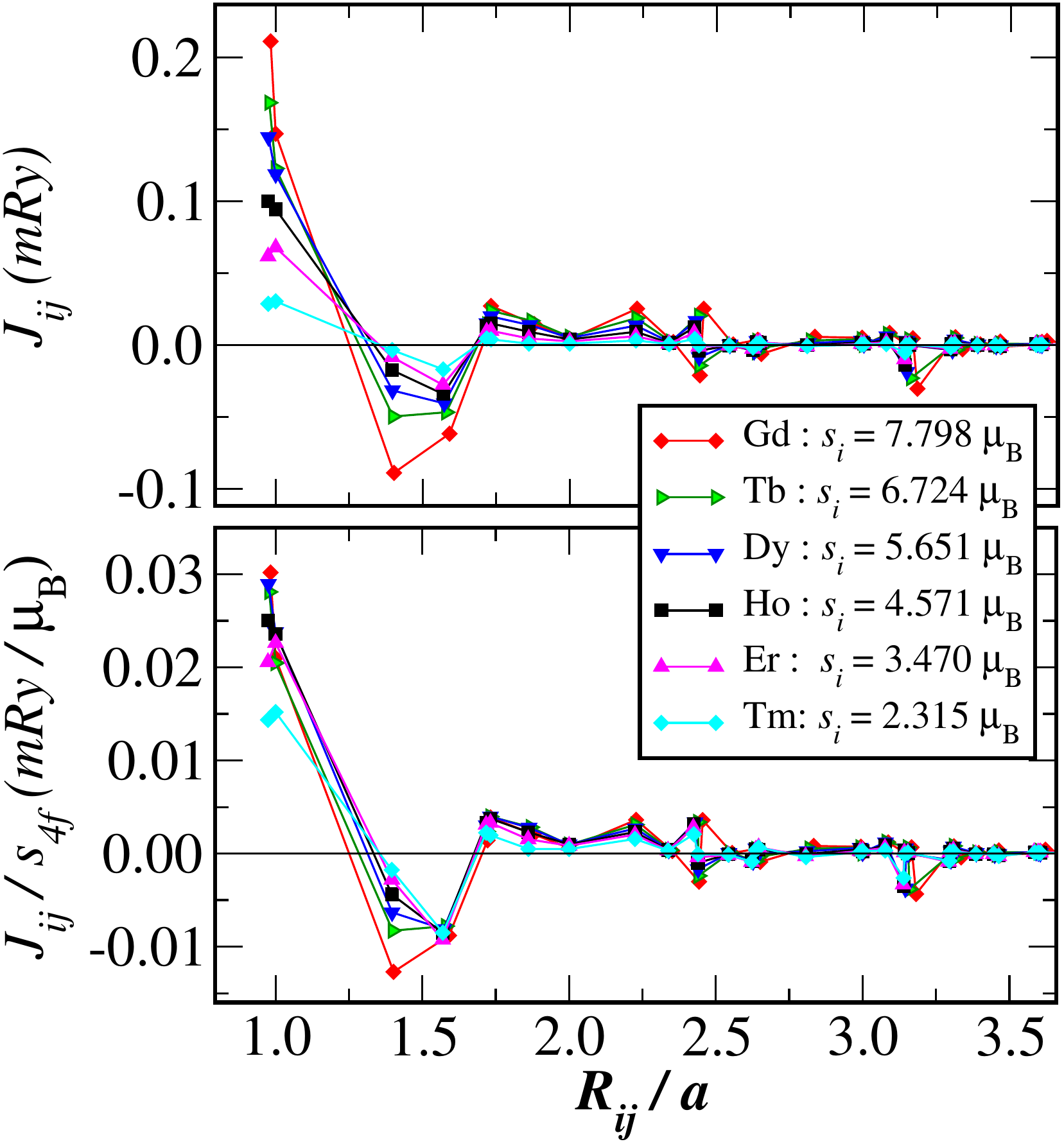} 
\caption{Upper panel: Calculated inter-site exchange parameters in heavy rare-earth metals as a function of the distance. $a$ denotes the lattice constant. Ferromagnetic reference state was assumed for all the elements.
Lower panel: $J_{ij}$-parameters divided by the value of the $4f$ spin moment for each element. For instance, in case of Gd: $S=7/2$.}
\label{RE-jijs}
\end{figure}

Next we have investigated the inter-site exchange interactions in the heavy REs, as defined in the first term of Eq.~\eqref{eq:SpinH}. We focussed on the heavy REs, since in the light REs, crystal field effects have a large impact. 
Figure~\ref{RE-jijs} shows the computed inter-site exchange interactions in heavy rare-earth metals. 
The parameters were extracted from the FM reference state. 
Note that this is the actual ground-state for Gd, Tb and Dy at low temperatures. 
The other elements have more complex non-collinear magnetic configurations~\cite{jensmac}.
Interestingly, in each coordination shell, the sign of the $J_{ij}$ is the same for all the considered elements. 
This is an indication that the underlying Fermi surfaces of all heavy lanthanides are very similar.
This fact has an experimental confirmation at least for Gd, Tb and Dy~\cite{REs-FS}.

For late REs it is known that the heavier the element, the smaller the unit cell volume.
Moreover, adding one more electron results in the decrease of the total spin moment, which is already anticipated from the Hund's rules picture.
This implies that there are two factors contributing to the changes in the $J_{ij}$ couplings along the heavy RE series: a decrease in the inter-atomic distances and in the spin moments. 
To disentangle these two effects, in the lower panel of Figure~\ref{RE-jijs} we show the re-scaled exchange parameters, divided by the value of the corresponding $4f$ spin moment (displayed in the legend of Figure~\ref{RE-jijs}).
One can see that the renormalized $J_{ij}$-parameters for all the considered elements with an exception of Tm are remarkably similar. This would suggest that the shape of the $J_{ij}$ curve is governed by the common features of the electronic structure of the REs e.g. the Fermi surface topology, while the strength of the exchange interaction is governed by the size of the induced moment of the valence states. A larger $4f$ spin-moment results in a larger induced valence moment (see Figure~\ref{fig:M}) and hence stronger interatomic exchange.
In fact, the present result implies that the Weiss field created by other neighbouring spins (i.e. the sum of all inter-site exchange interactions) is close to be constant throughout the series of the heavy REs.

As stated before, we extract the exchange parameters from a ferromagnetic state. However it is well known that some of the REs have a more complicated magnetic ordering~\cite{jensmac}. We will discuss the stability of the ferromagnetic state in more detail below.

\subsubsection{Ordering temperature}

\begin{figure}
\centering
\includegraphics[width=\columnwidth]{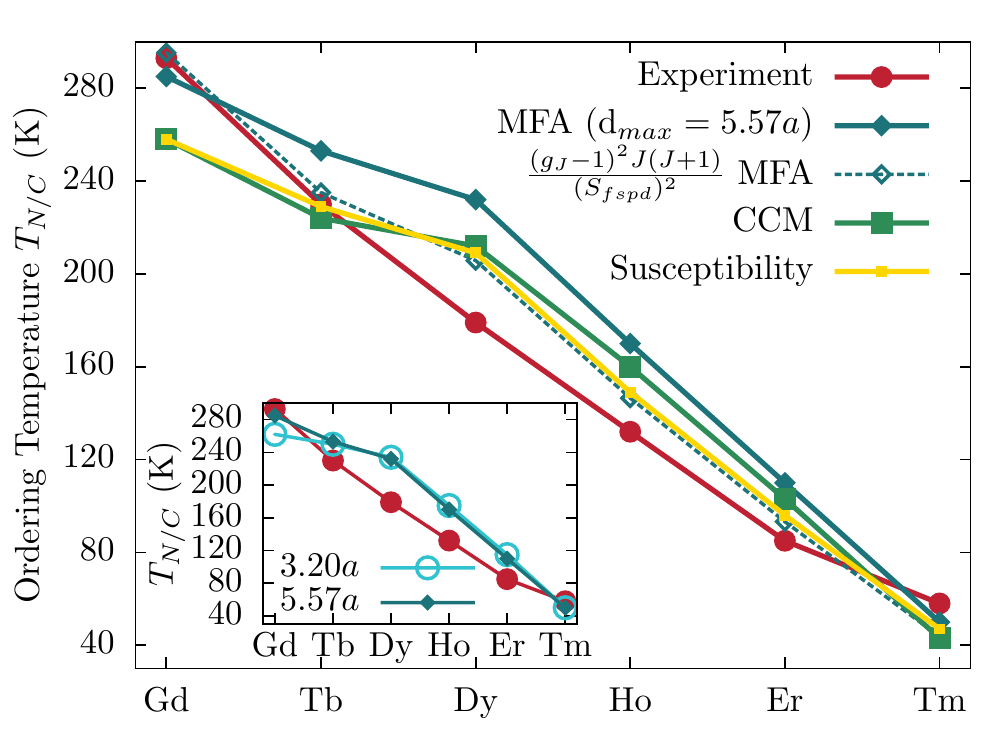}\\  
\caption{\label{fig:Tc} Calculated $T_{N/C}$ for heavy REs. The results are obtained using $J_{ij}$'s presented in Figure~\ref{RE-jijs}. The experimental data for the ordering temperature, \textit{i.e.} the transition from a paramagnetic state to an ordered state, is taken from Ref.~\cite{jensmac}. These ordered states have no net magnetic moment, with an exception for Gd, where the transition is directly to the ferromagnetic state. The solid lines are calculated with the classical spin moment, the dashed line is calculated with the de Gennes prescription and is merely a rescaling of the classical MFA result.}
\end{figure}

In Figure~\ref{fig:Tc} we show the calculated ordering temperatures ($T_{N/C}$) for the series of heavy REs.
The results were obtained with Monte Carlo and MFA methods, utilizing the $J_{ij}$'s presented in Figure~\ref{RE-jijs} as defined in the first term of Eq.~\eqref{eq:SpinH}. In all simulations the total spin moment (see legend of Figure~\ref{RE-jijs}), including $f$ and $[spd]$ contributions, was used in the classical limit ($(S_{fspd}^{z})^2$). Additionally we plot the ordering temperatures in the MFA when de Gennes prescription $S^2 \rightarrow (g_J-1)^2J(J+1)$ is used. In this case, the total angular momentum quantum number $J$ is only due to the $f$ electrons. Although both approaches rely on the same physical mechanism, namely that the exchange interaction is between spin moments, de Gennes prescription takes into account the existence of an orbital moment, by taking the projection of the spin onto the $J$-axis.

We have also verified the impact of varying the amount of neighbour interactions on the resulting $T_{N/C}$ values.
For this study we used the computationally less demanding MFA-based estimation.
In the inset of Figure~\ref{fig:Tc} the results obtained with two different cut-off radii for the $J_{ij}$ interactions are compared.
One can see that an increase of the cluster radius from 3.2$a$ to 5.57$a$ produces almost negligible changes in the calculated ordering temperature for almost all heavy REs.
The exceptional case is Gd, where the differences are more significant, as was also shown in Ref.~\cite{0953-8984-15-17-327}.
We believe that it is related to the fact that the strength of the $J_{ij}$'s in this metal is the largest among the studied systems. The $T_{N/C}$'s obtained from the maximum in the susceptibility are calculated with a cut-off radius of  5.57$a$. For the CCM we used only 3.2$a$, except for Gd where we used 5.57$a$.

From Figure~\ref{fig:Tc} one can see that both Monte Carlo methods ($T_{N/C}$ obtained from the maximum in the susceptibility and using CCM) and the MFA produce quite similar results.
MFA has a well-known tendency to over-estimate the $T_{N/C}$'s. 
Indeed, our results indicate that MFA-based estimates are about 20\% larger than the ones obtained with Monte Carlo methods.
One can also see that the calculated $T_{N/C}$'s produced with both methods are in fair agreement with experimental data.
The largest errors in the Monte Carlo calculation, are found for Gd and Dy and reach about 35 K.
However, in spite of these differences, the qualitative trend of lowering of the $T_{N/C}$ across the heavy RE series is nicely reproduced in these calculations.

A small remark on the experimental $T_{N/C}$ in Figure~\ref{fig:Tc}: We chose to compare our results to the experimental ordering temperature related to magnetically-ordered to paramagnetic transition. However, for most heavy rare earth elements there is a low temperature ferromagnetism, followed by a phase without net moment (helix, cone or longitudinal spin-wave) before the paramagnetic phase. In principle Monte Carlo simulations should be able to reproduce both transitions. However, to investigate the full magnetic phase diagram one needs both temperature dependent $J_{ij}$'s and a temperature dependent anisotropy. These quantities, we can not calculate at the moment.

\subsubsection{Fourier transform exchange parameters\label{res:mag}}

\begin{figure}[]
\includegraphics[width=\linewidth]{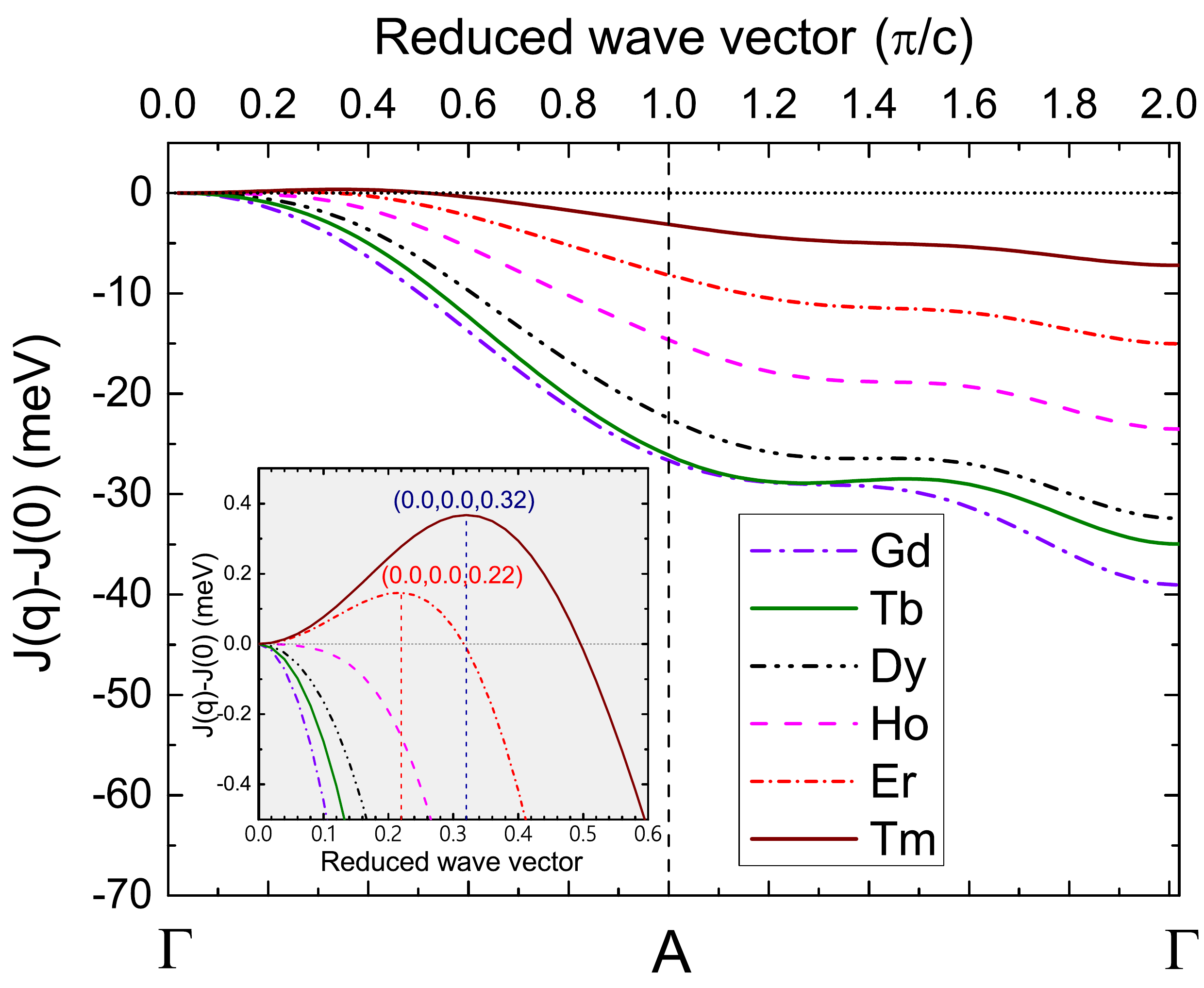}
\caption{\label{fig:Jq} Fourier transform of the exchange interaction $J(q)-J(0)$ for heavy rare-earth metals plotted along the $\Gamma-A-\Gamma$ line. The inset shows a magnification of the figure for the reduced wave vector in the interval [0, 0.6]. In the inset we also indicated the pitch vector for Er and Tm, showing that the ferromagnetic reference state is unstable for both metals. }
\end{figure} 

The Fourier transformed exchange constants $J(q)$ shifted by the value at the $\Gamma$ point, are shown in Fig.~\ref{fig:Jq} for heavy rare-earth metals. The values have been calculated by using linear spin-wave theory in the framework of the adiabatic approximation and are plotted in the reciprocal space along the path $\Gamma-A-\Gamma$, with $A-\Gamma$ laying in the second Brillouin zone. The reported results for Gd, Tb, Dy and Ho have been computed by using as a reference state a collinear configuration with the magnetic moments pointing parallel to the basal plane. The positive maximum at finite $\mathbf{q}$ vector for Er and Tm indicates that for these elements the ferromagnetic state is unstable by about 0.15 and 0.36 meV, respectively. In experiments at low temperature, indeed a spin spiral is found for Er and a longitudinal collinear spin-wave for Tm~\cite{jensmac}.  
The maxima in $J(q)-J(0)$ occur at pitch vectors (0,0,0.22) and (0,0,0.32) in $\pi/c$ units, that are different from the experimental data in Ref.~\cite{PhysRevB.61.4070}. This might be related to the fact that the $J_{ij}$?s were extracted not from the actual ground state, but from the FM state. The low temperature experimental magnetic ground states for Gd, Tb and Dy are ferromagnetic~\cite{jensmac}, which is in agreement with our findings. Experimentally Ho is found to have a spin-spiral ground state, whereas we find a stable ferromagnetic ground state. In Ref.~\cite{0295-5075-49-6-775}, the authors also found a ferromagnetic ground state for Ho, however, it was found to be unstable for slightly reduced $4f$ moments. 
Overall our results give encouraging agreement with experiment, although there is room for improvement. In order to reproduce the full magnetic phase diagram of these elements one needs to calculate theoretically the magnetic anisotropy and exchange interactions as a function of temperature, something which can not be done routinely with HIA. Also, in order to evaluate the full details of the low temperature magnetic phase of the rare-earths one needs in general to also include a full description of the crystal field effects. This is outside the scope of the present study, although in principle the HIA should be a relevant starting point for such an investigation, and offer a better ansatz than for instance the LDA+U or 4f as core approximations.


\subsubsection{Magnon spectra\label{res:mag}}
\begin{figure}[b]
     \subfigure[Spin wave dispersion spectrum.\label{fig:sqw-T78K-23NN}]{
		\includegraphics[width=\columnwidth]{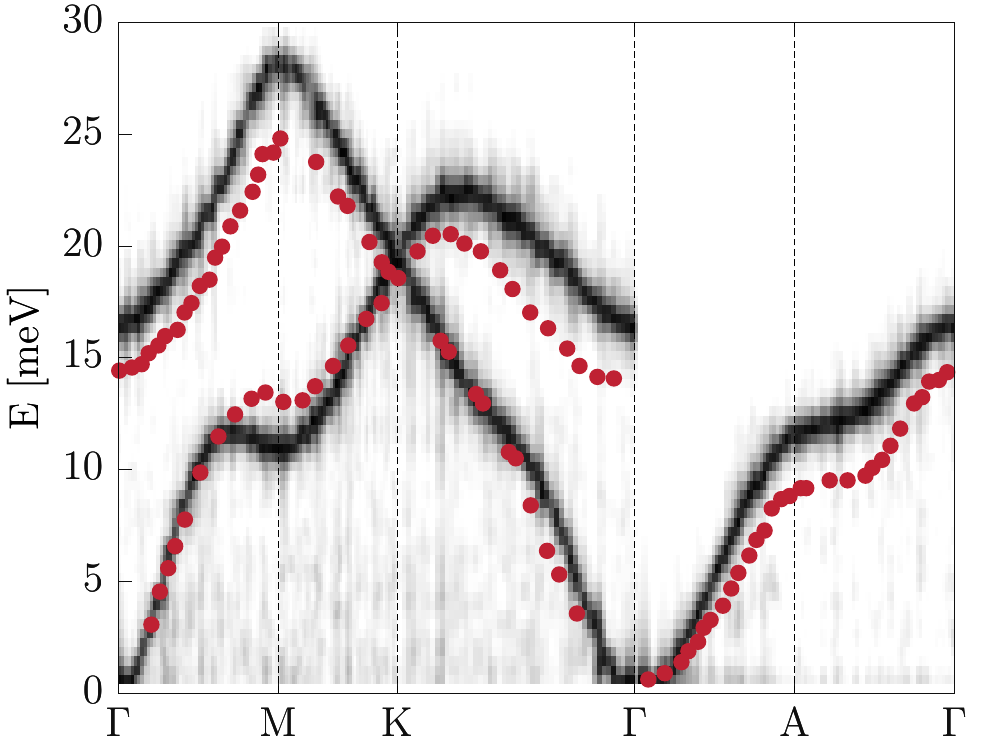}}  
     \subfigure[Adiabatic magnon spectrum.\label{fig:adiabatic}]{
		\includegraphics[width=0.97\columnwidth]{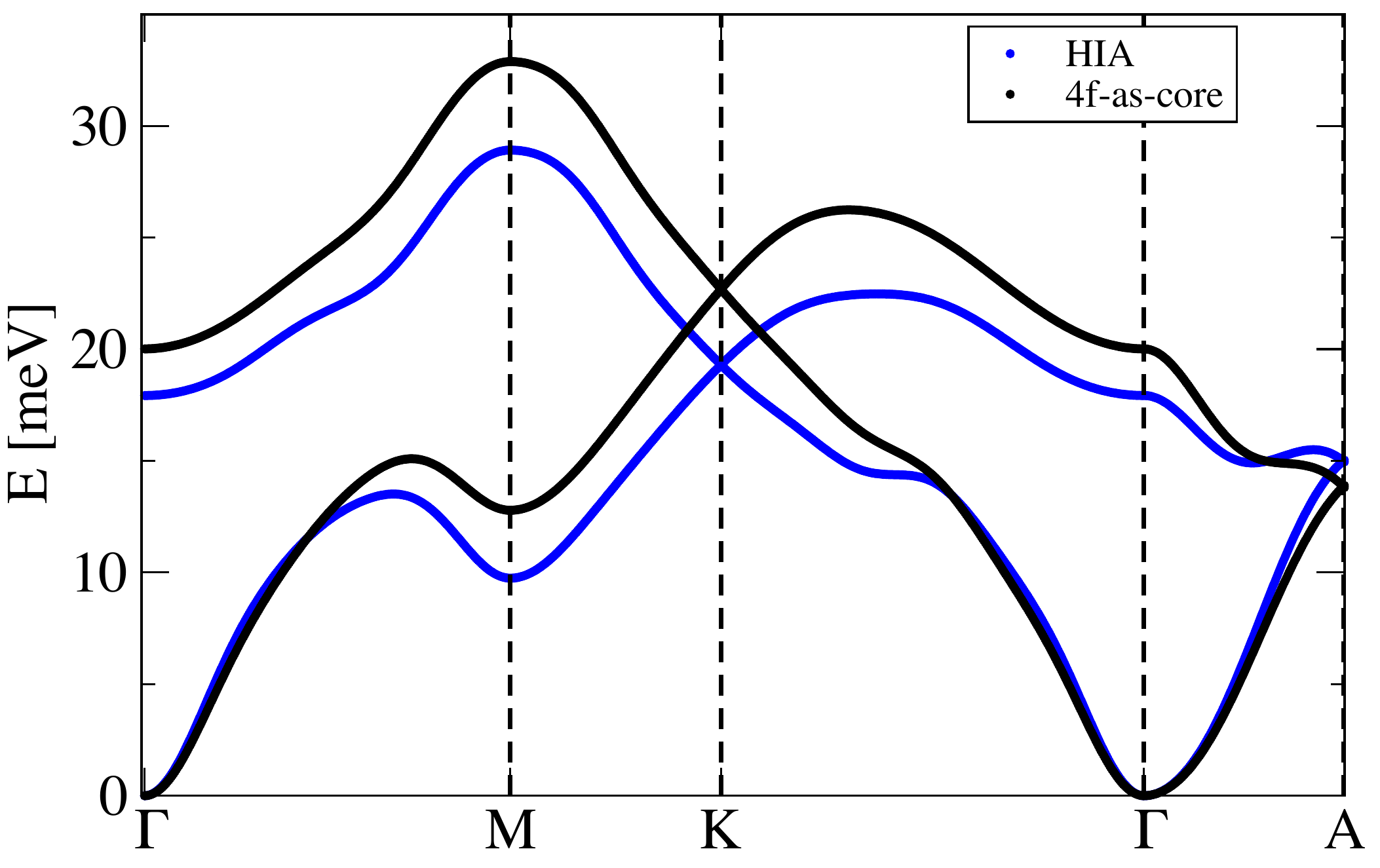}} 
\caption{\label{fig:Gd-magnons} Spin wave dispersion spectrum of hcp Gd.  (a) Simulated spectrum using ASD (black) along with experimental data (red filled circles) from Ref.~\cite{jensmac}. (b): A comparison between adiabatic magnon spectra calculated with exchange parameters obtained with HIA (blue) and with $4f$-as-core (black).}
\end{figure}

As an example we show in Figure~\ref{fig:sqw-T78K-23NN} the calculated magnon spectra of Gd (black) obtained with ASD using the exchange parameters in Figure~\ref{RE-jijs}. This element was selected because its magnon spectrum is dominated by exchange effects, in contrast to many of the other rare-earths where either crystal field effects and magnetic anisotropy become important, or where even magnon-phonon coupling is important~\cite{jensmac}.
First of all, one can see in Figure~\ref{fig:sqw-T78K-23NN}, that for small $q$ values, the dispersion follows a parabolic dependence. 
This is a clear evidence of the FM ordering, intrinsic to Gd at low temperatures. 
Second, both acoustic and optical branches are in good agreement with the experimental data, indicating the high quality of the obtained parameterization of the Heisenberg model from DFT for Gd, as well as the accuracy in extracting the dynamical structure factor from spin-dynamics simulations.\\

Finally we compare in Fig~\ref{fig:adiabatic} the adiabatic magnon spectra obtained using the exchange parameters calculated with two different methods compatible with the standard model, namely $4f$-as-core and HIA (for more details, see Appendix~\ref{app:cscHIA}). One can see that the overall shape of the two magnon curves is rather similar throughout the whole Brillouin zone and the spectra differ by not more than $4$~meV. The similarity of these two results provides an other justification for the use of the $4f$-as-core method to simulate the dynamical magnetic properties of the REs.

\subsection{Spectral properties\label{sec:spectra}}

\begin{figure*}
     \centering
     \subfigure{\includegraphics[width=.32\textwidth]{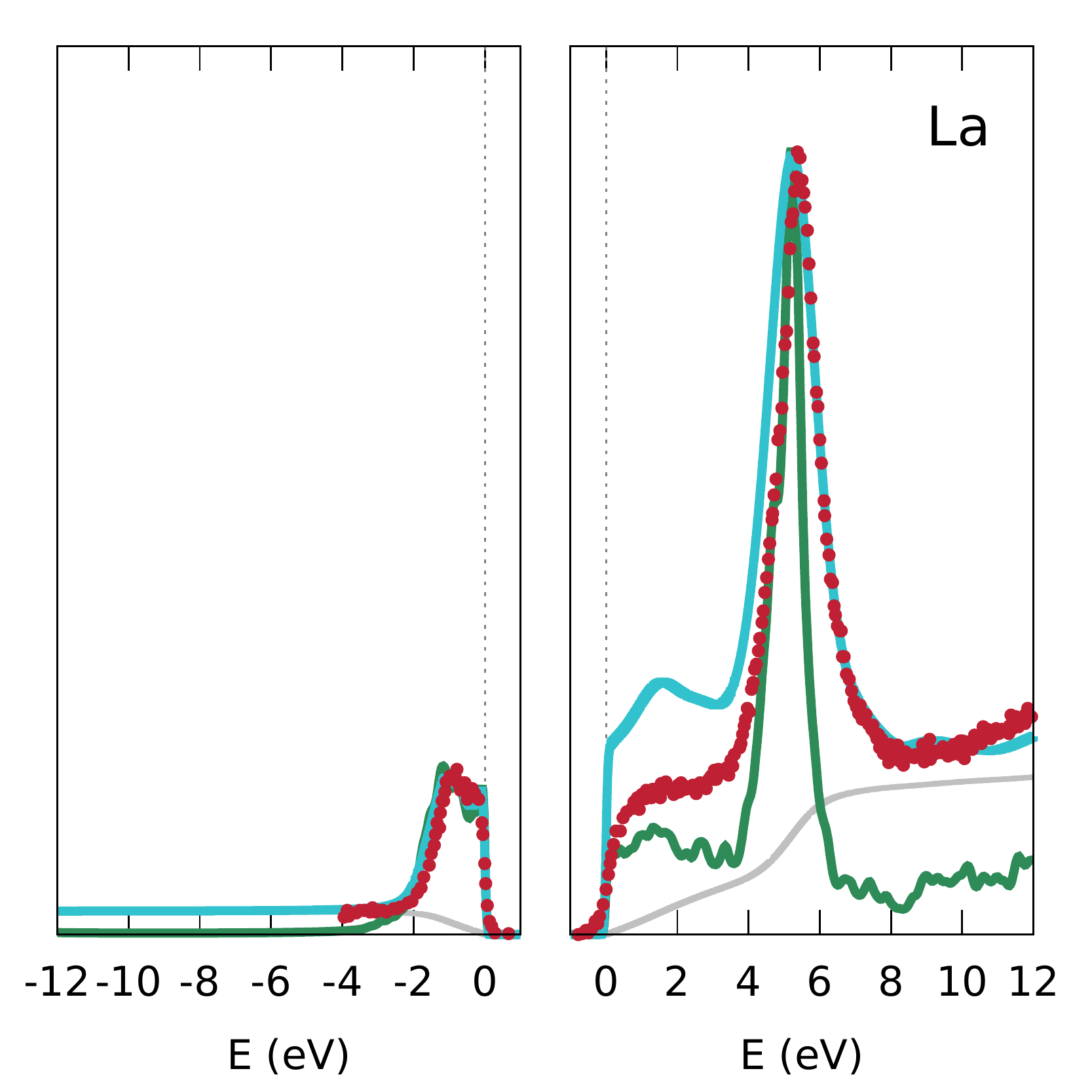}} 
     \subfigure{\includegraphics[width=.32\textwidth]{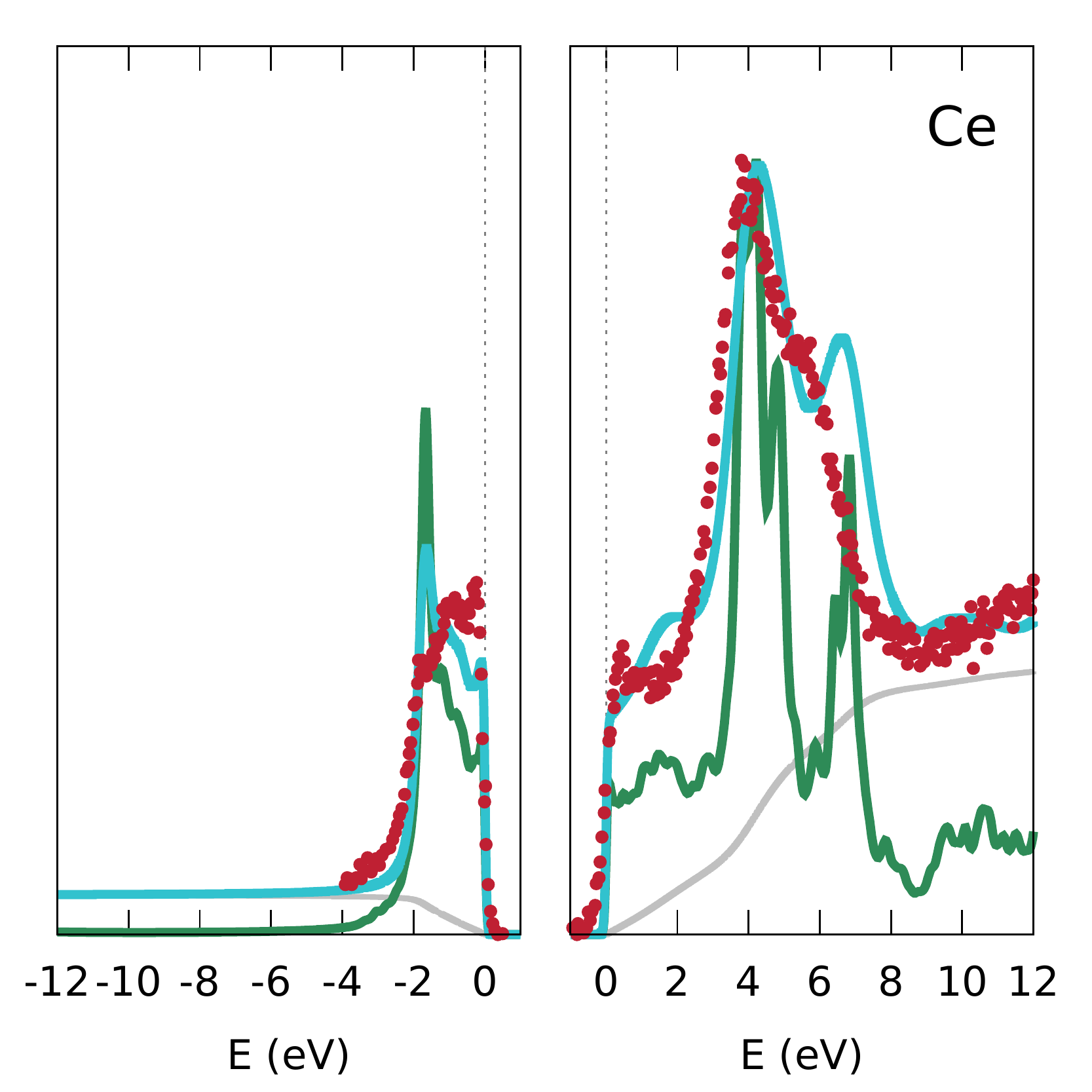}} 
     \subfigure{\includegraphics[width=.32\textwidth]{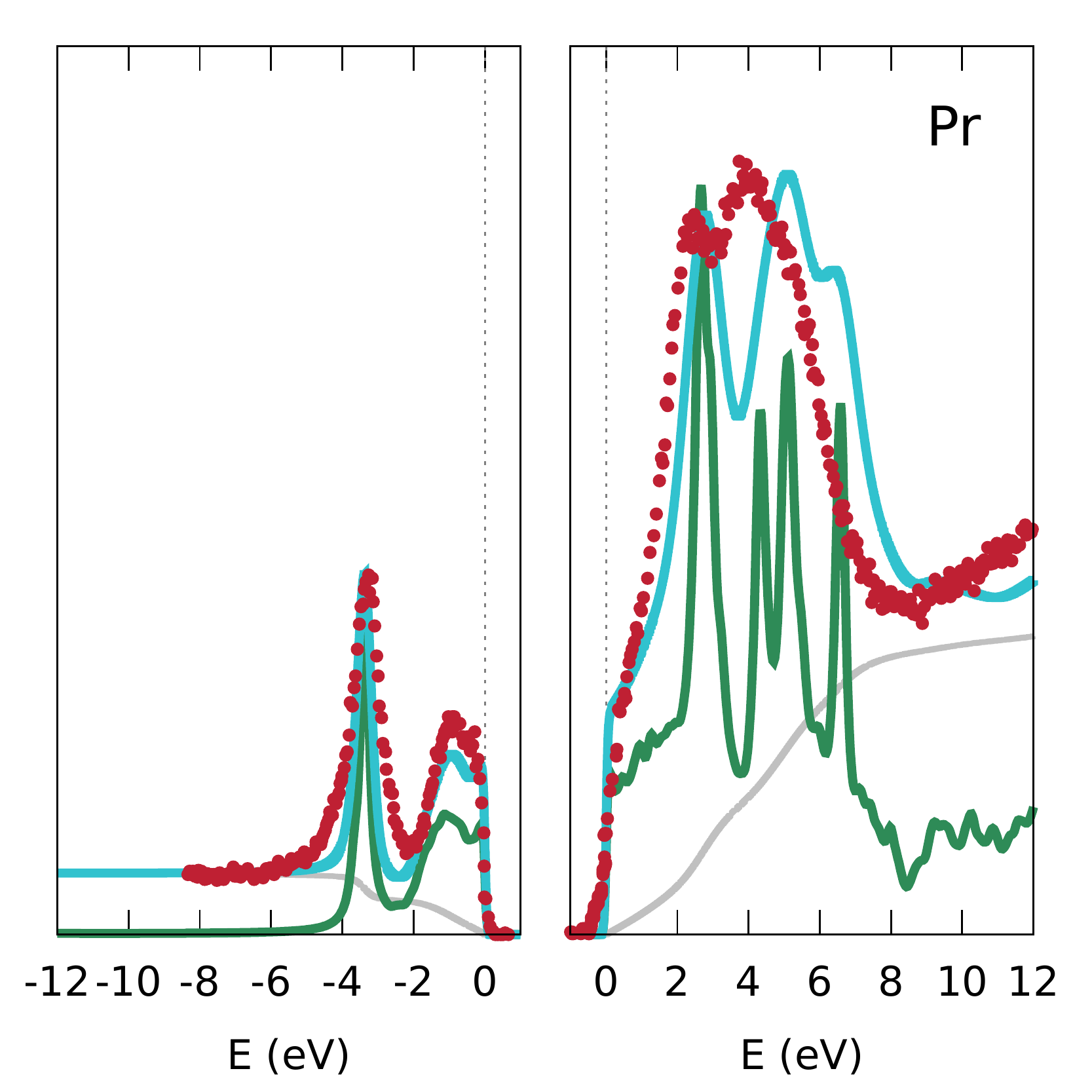}} 
     \subfigure{\includegraphics[width=.32\textwidth]{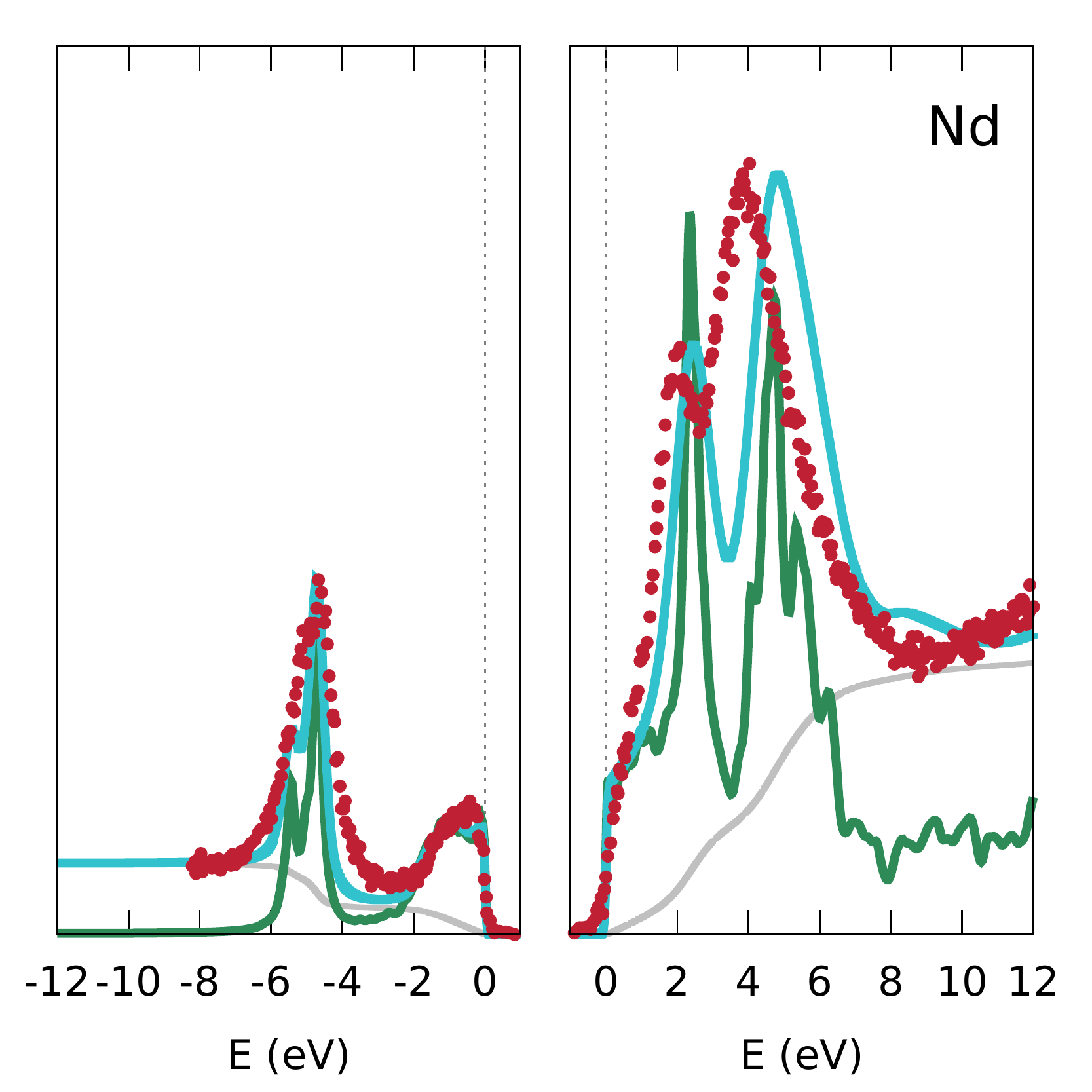}} 
      \subfigure{\includegraphics[width=.32\textwidth]{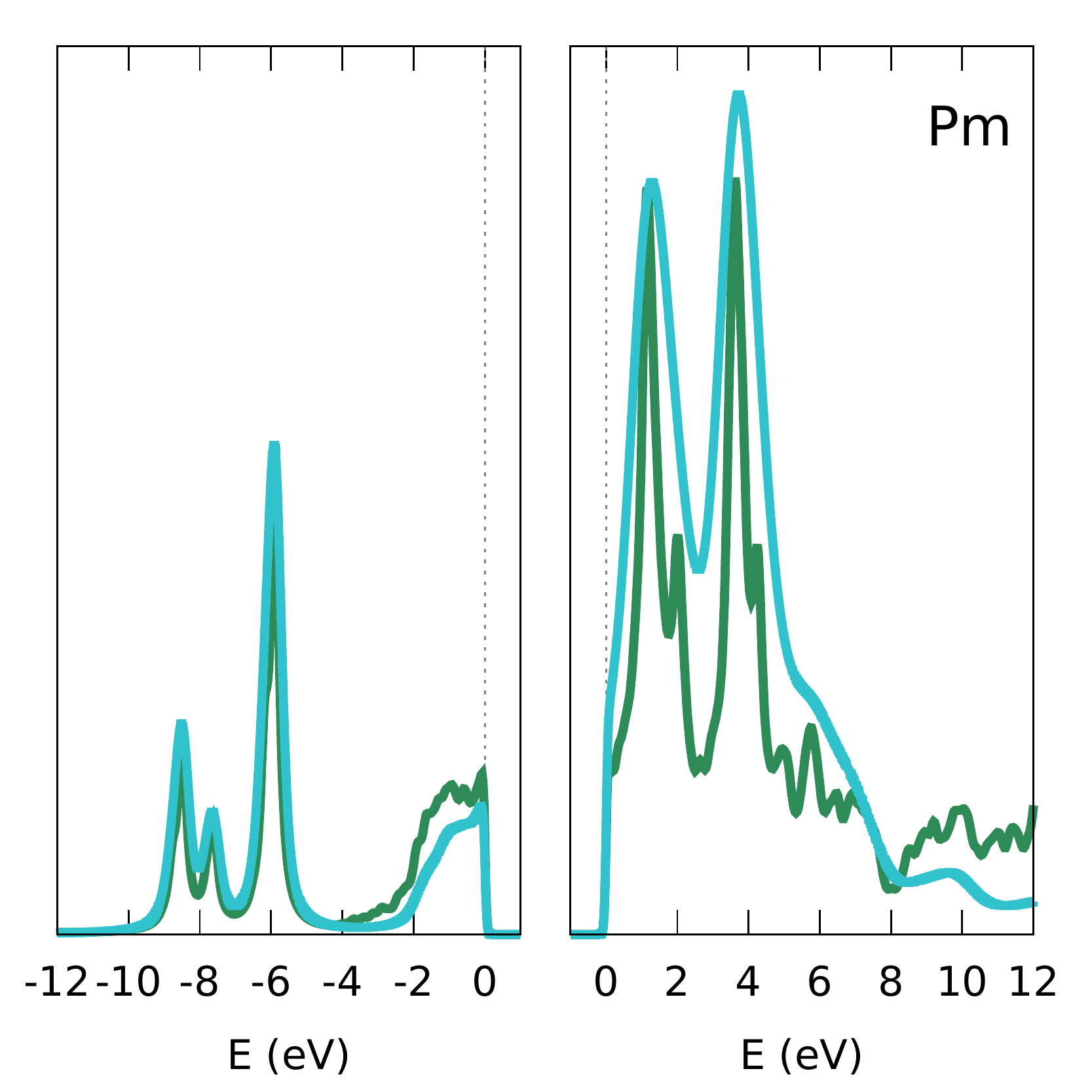}} 
      \subfigure{\includegraphics[width=.32\textwidth]{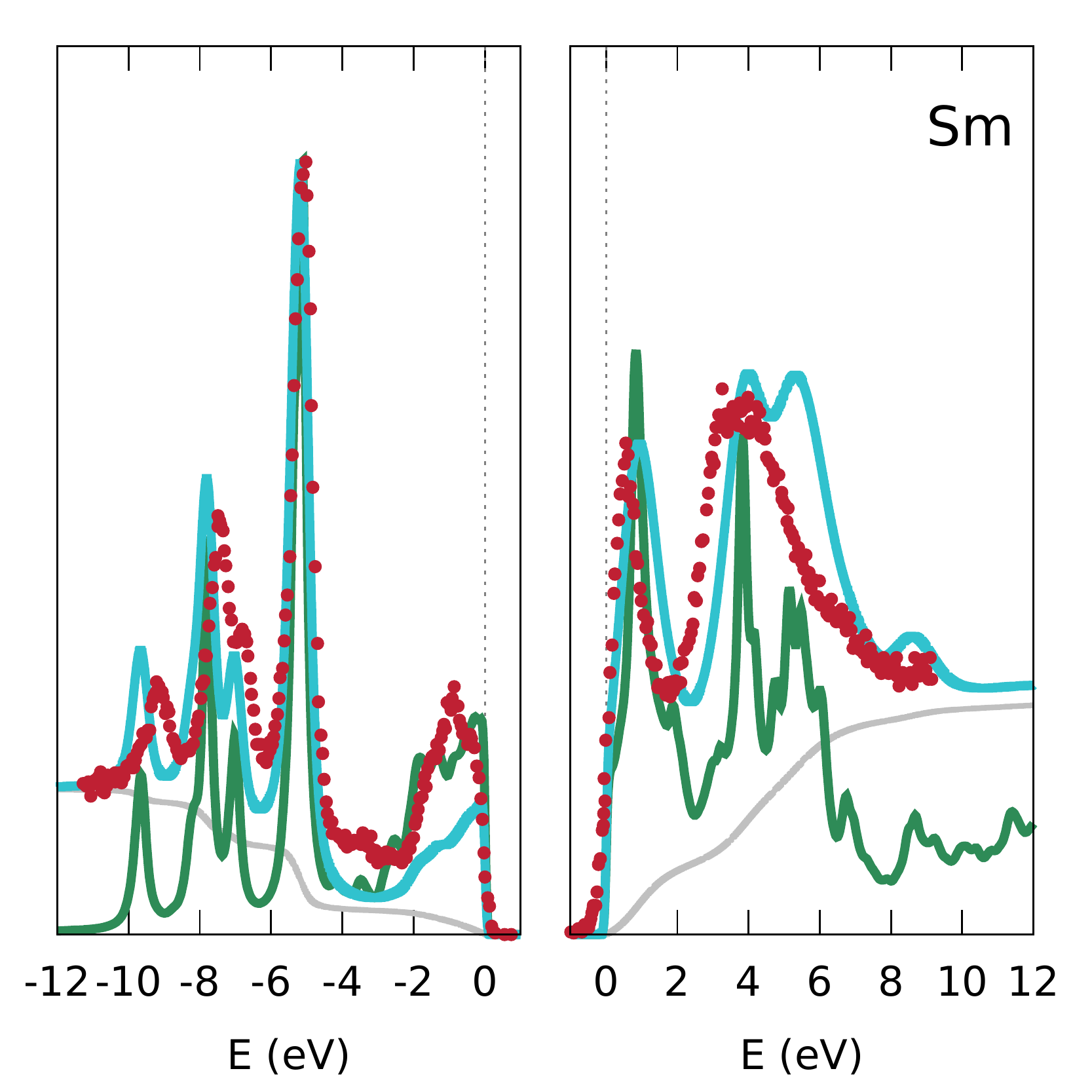}} 
     \subfigure{\includegraphics[width=.32\textwidth]{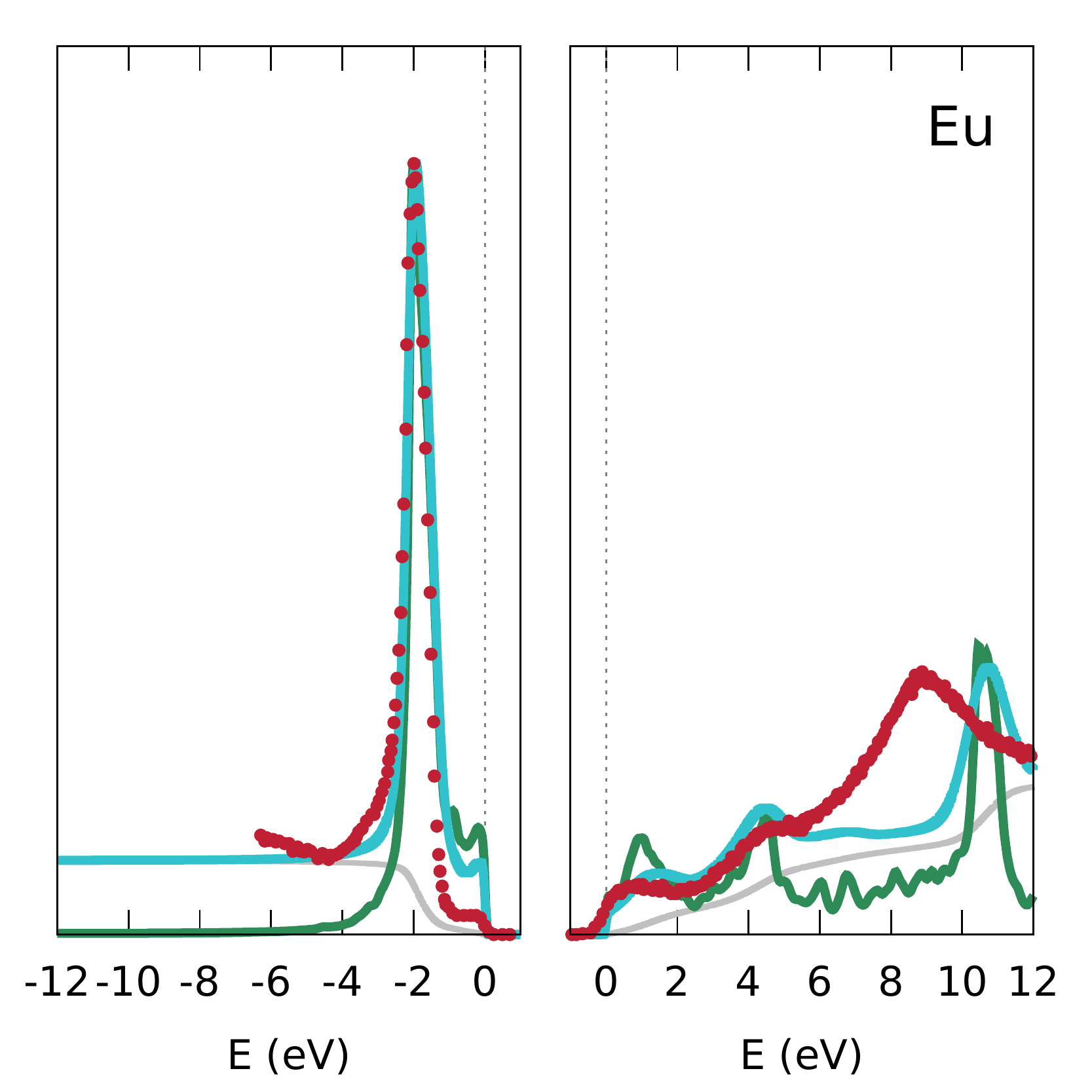}} 
      \subfigure{\includegraphics[width=.32\textwidth]{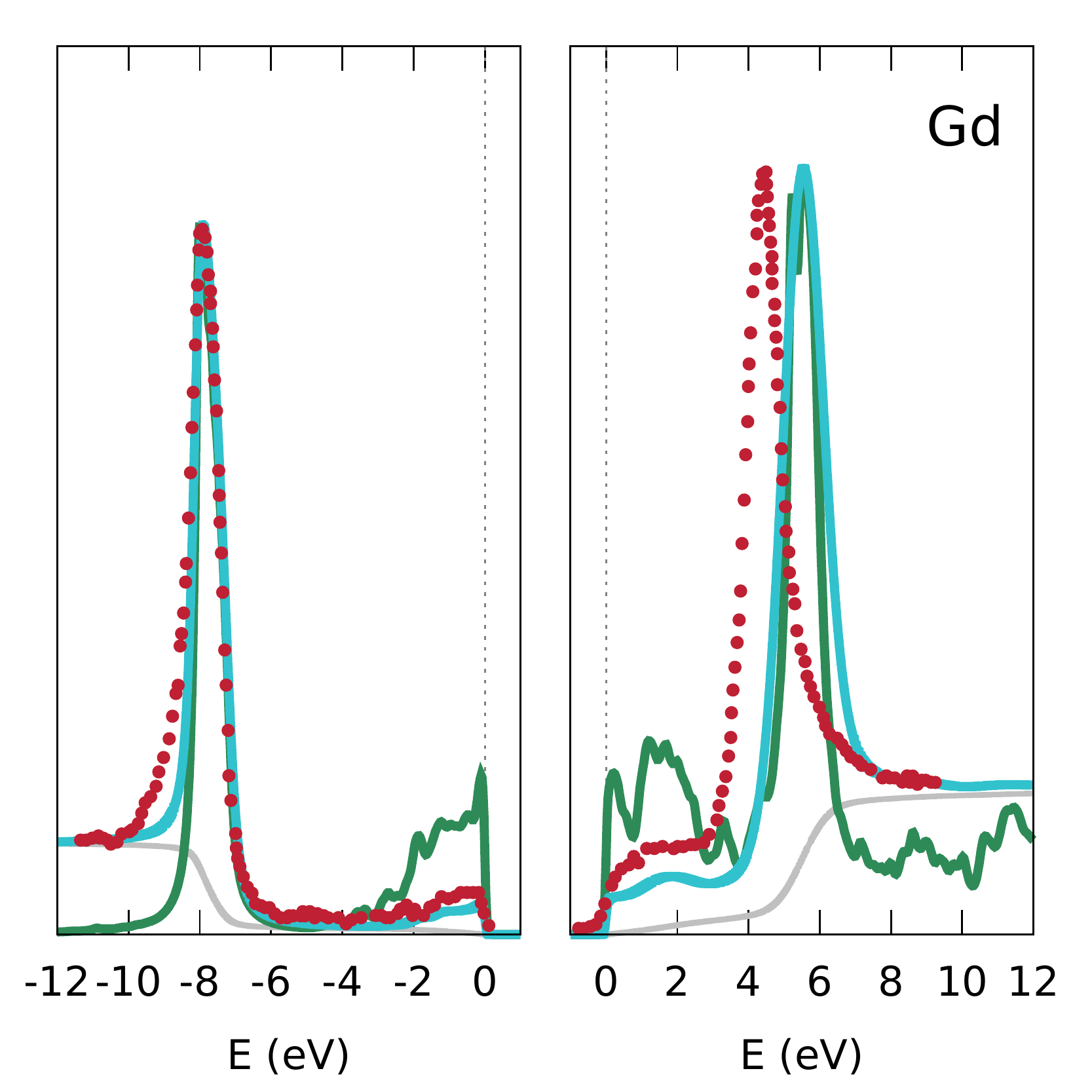}} 
      \subfigure{\includegraphics[width=.32\textwidth]{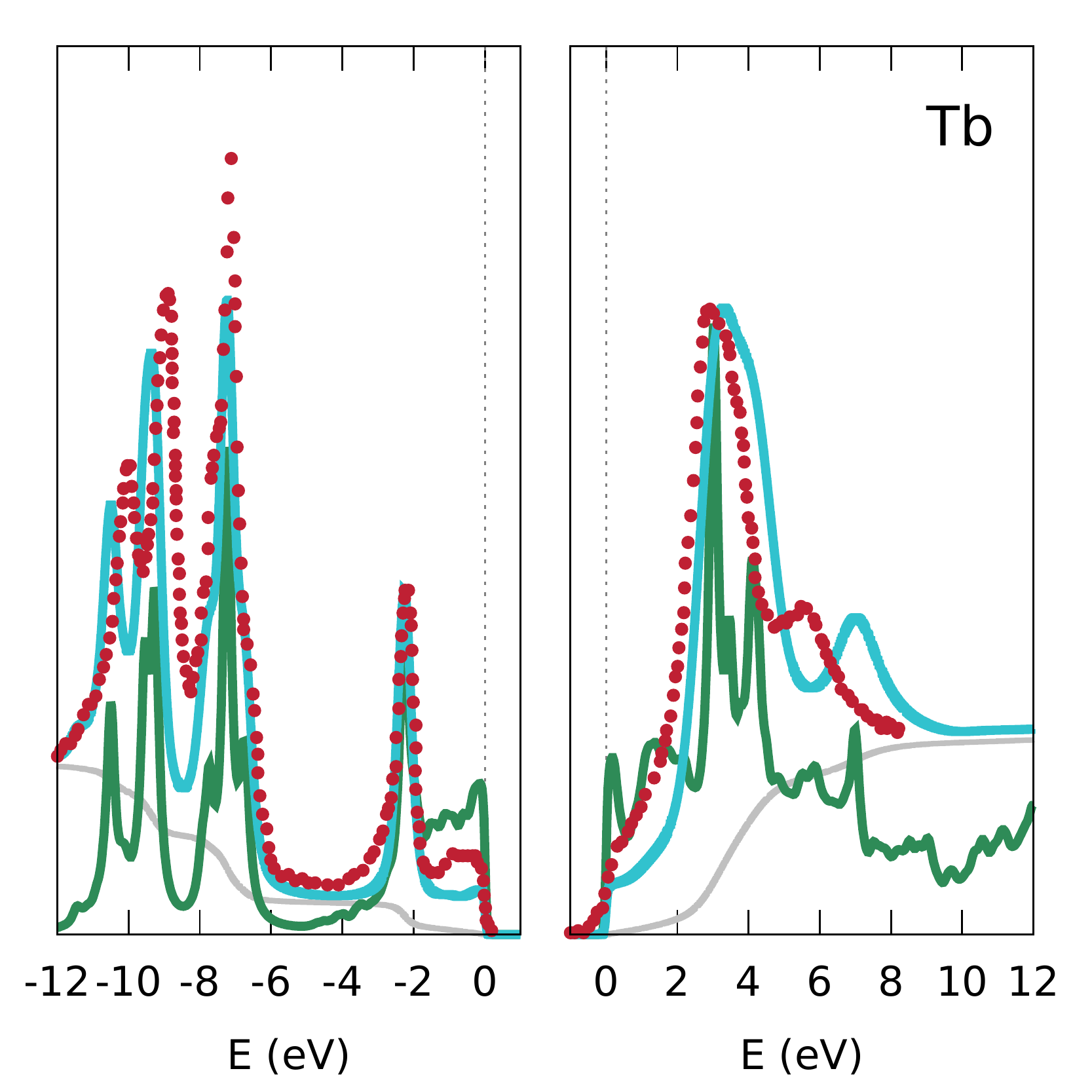}} 
\caption{Spectra: The red dots are experimental data from Ref.~\cite{LangCox1981}. The green line is the bare total spectral function calculated in the Hubbard I approximation. For the blue line we applied a Gaussian smearing of FWHM=0.3eV for XPS and FWHM=1.0eV for BIS, we multiplied the $f$ and $d$-partial density of states by their cross-sections just below (XPS) and just above (BIS) the Fermi level and we added a background proportional to the integral under the spectrum. For  clarity we also plotted this background in grey. For each element the plot consists of the XPS part (left) and BIS part (right). Since Pm has a radioactive unstable nucleus, no experimental data are available. Therefore the experimental data are not included in the plot and the blue line does not contain a background. For the same reasons, we used the fully localized limit double counting scheme for this particular element.  \label{fig:spectra1}}
\end{figure*}

\begin{figure*}
     \centering
     \subfigure{\includegraphics[width=.32\textwidth]{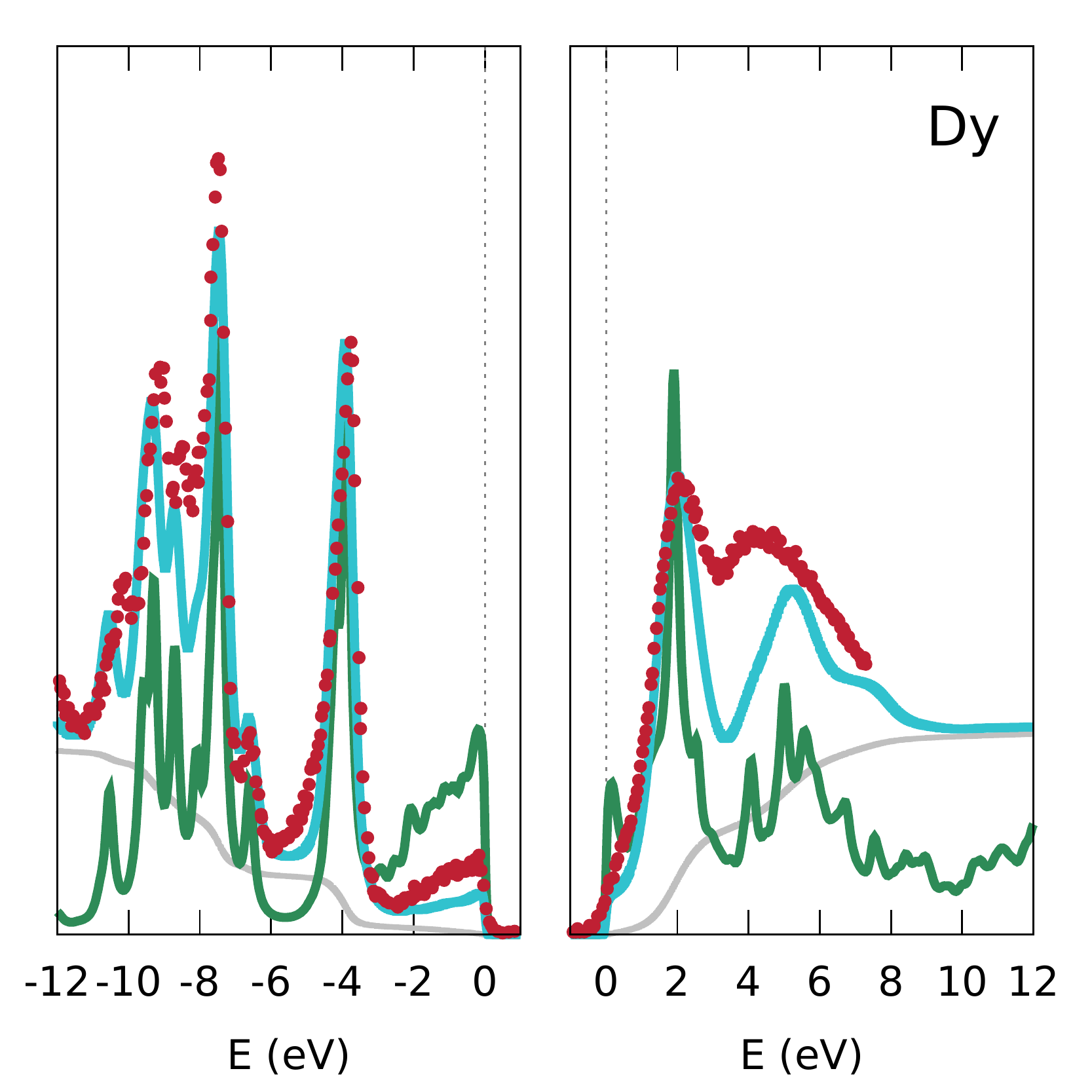}} 
      \subfigure{\includegraphics[width=.32\textwidth]{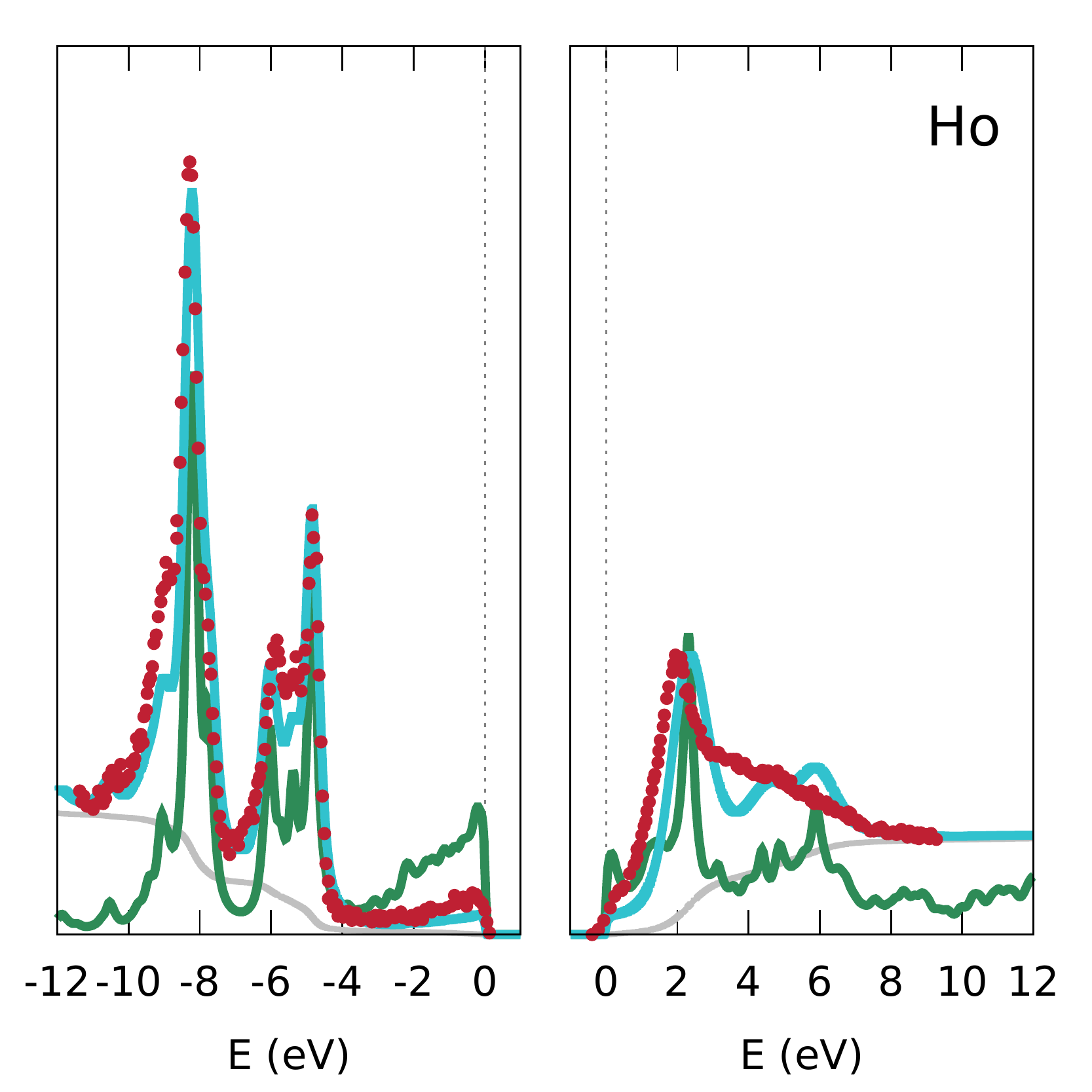}} 
      \subfigure{\includegraphics[width=.32\textwidth]{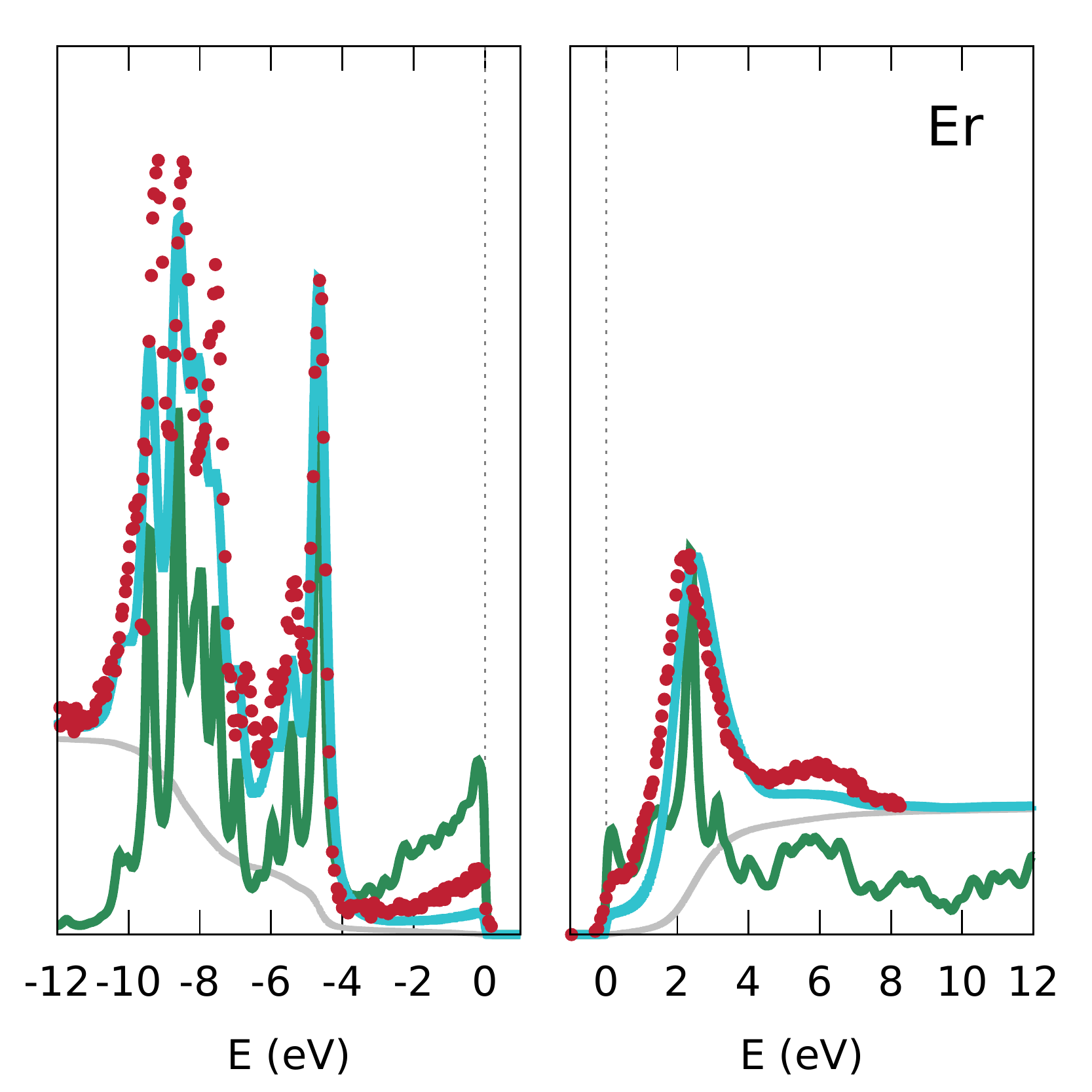}} 
     \subfigure{\includegraphics[width=.32\textwidth]{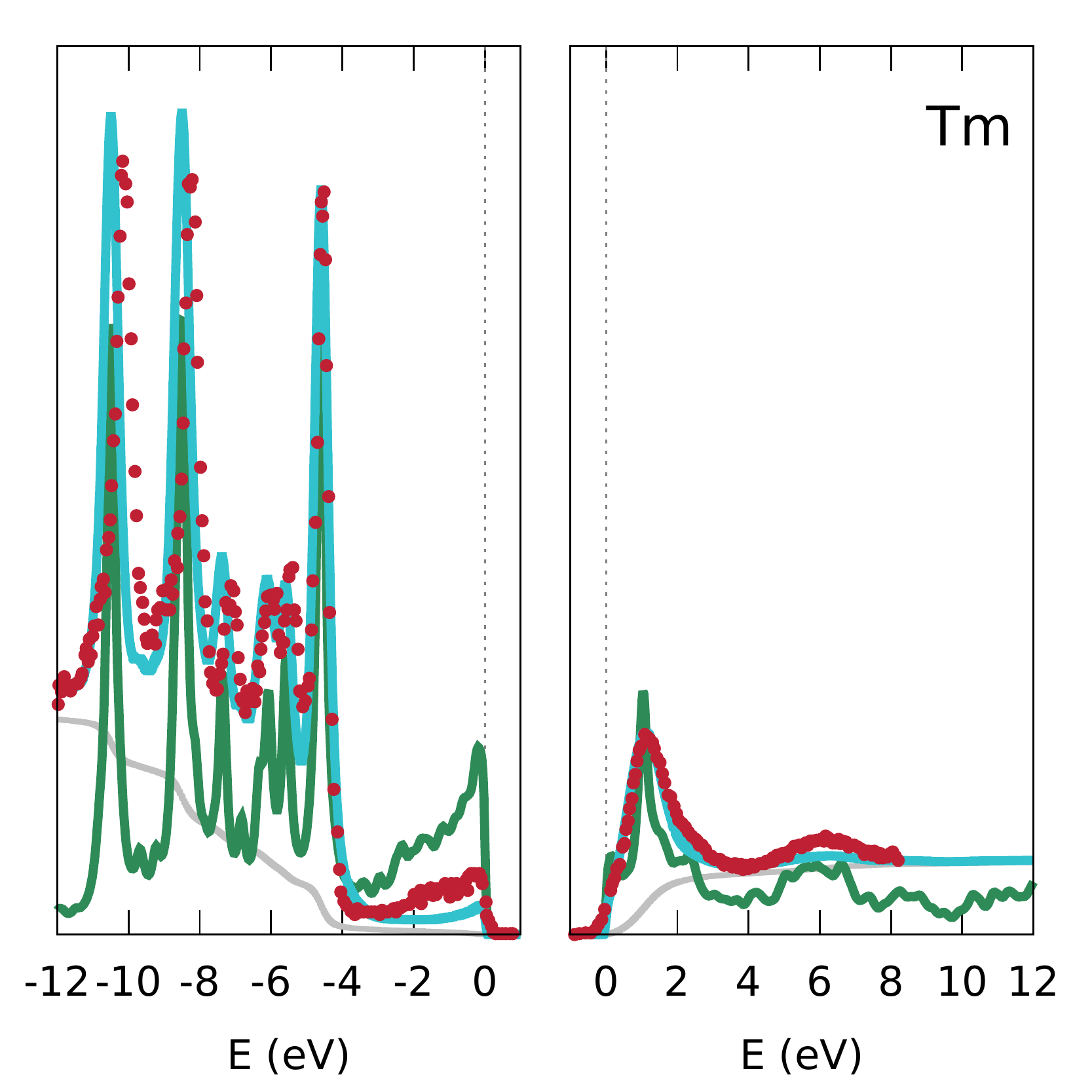}} 
      \subfigure{\includegraphics[width=.32\textwidth]{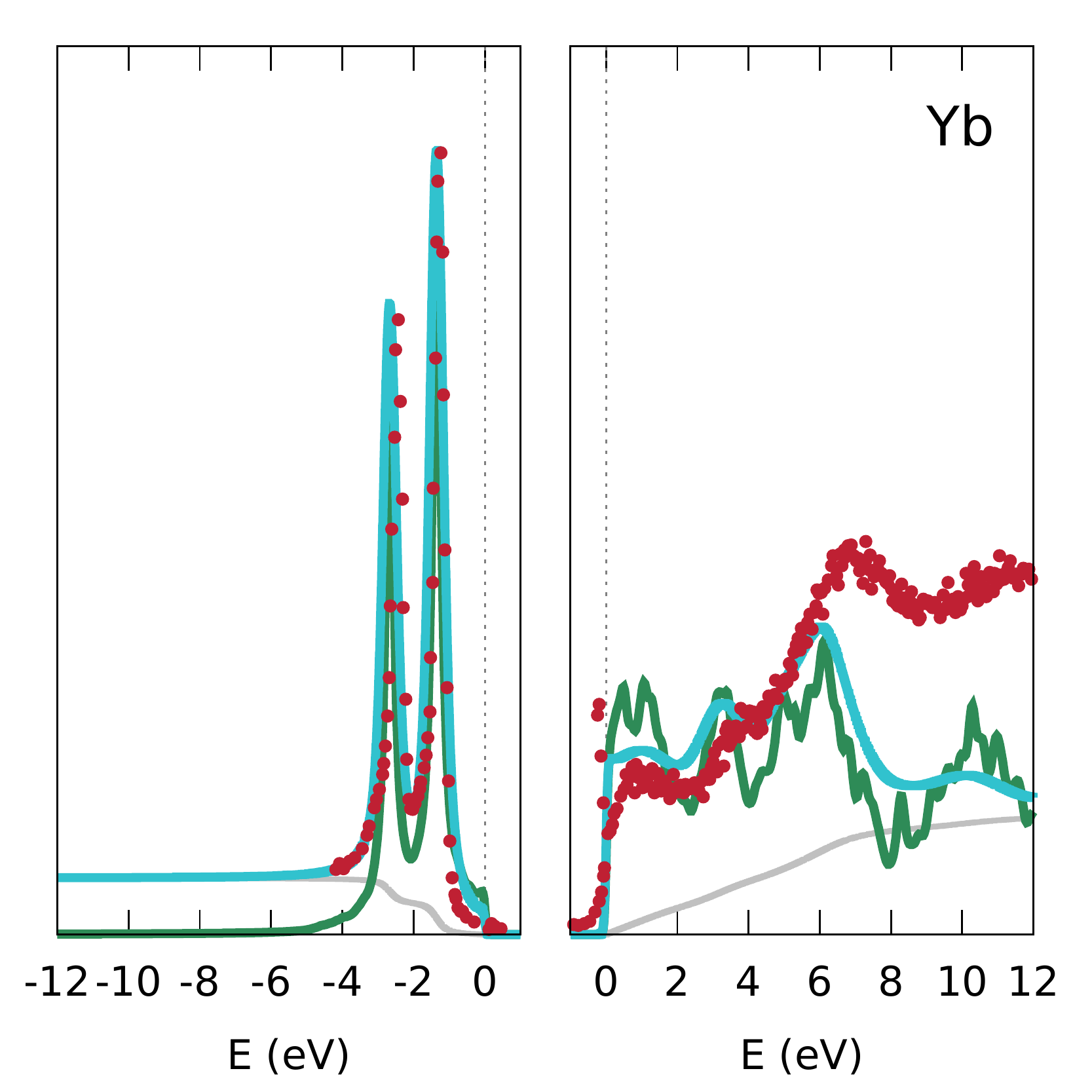}} 
      \subfigure{\includegraphics[width=.32\textwidth]{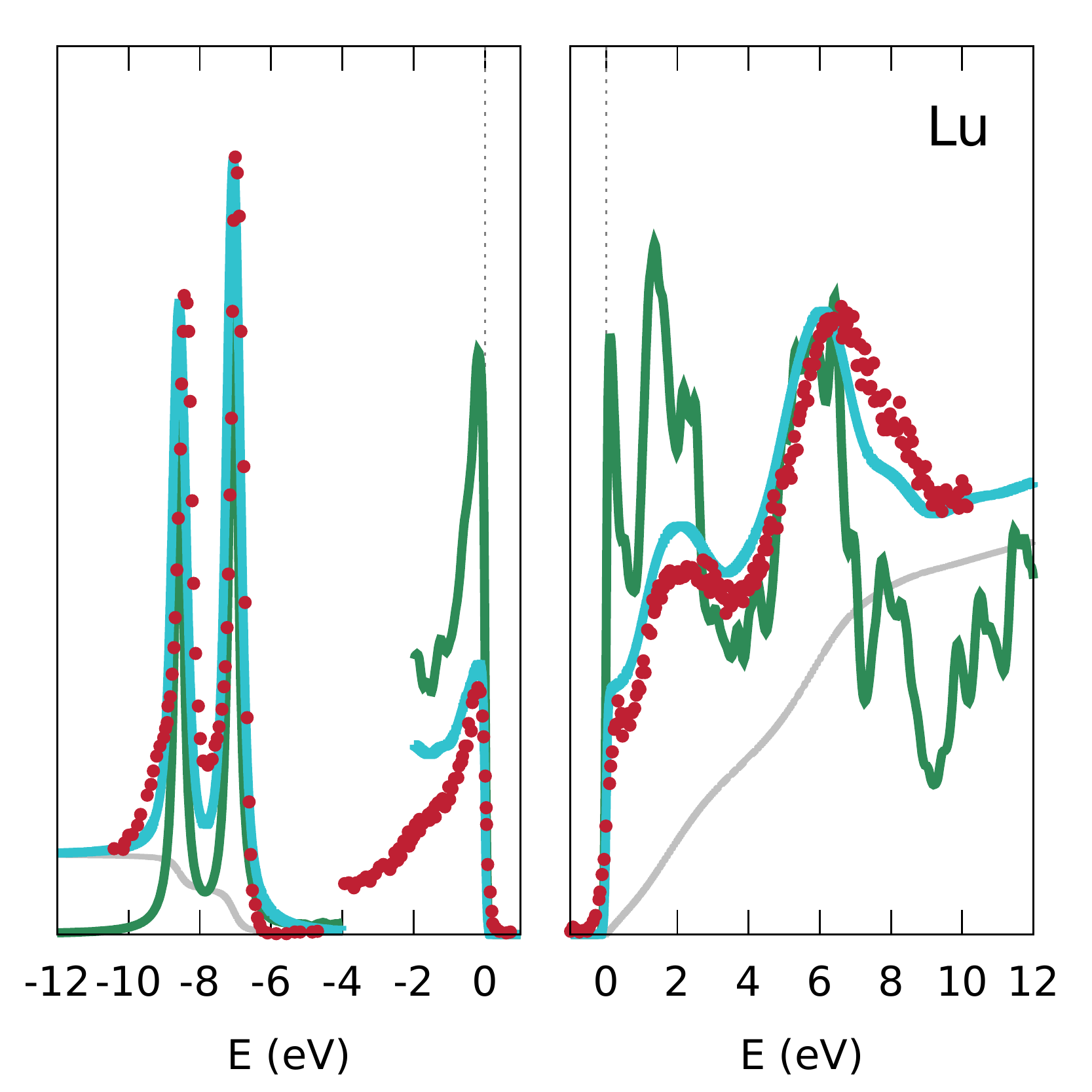}} 
\caption{Spectra: The red dots are experimental data from Ref.~\cite{LangCox1981}. The green line is the total spectral function calculated in the Hubbard I approximation. For the blue line we applied a Gaussian smearing of FWHM=0.3eV for XPS and FWHM=1.0eV for BIS, we multiplied the $f$ and $d$-partial density of states by their cross-sections just below (XPS) and just above (BIS) the Fermi level and we added a background proportional to the integral under the spectrum. For  clarity we also plotted this background in grey. For each element the plot consists of the XPS part (left) and BIS part (right).  Note that for Lu the spectrum in the XPS panel contains just below the Fermi level the spectral function with the same scaling as the BIS spectral function, which is different from the scaling of the XPS spectral function between -4 and -12 eV. \label{fig:spectra2}}
\end{figure*}

The photoemission spectra were calculated at the experimental volumes. An LDA+HIA functional was used for the plots shown here, but GGA+HIA was found to give very similar results. 
As discussed in Sec.~\ref{sec:compdet_coh}, the double counting correction for the spectra was chosen such that the position of the first occupied or unoccupied peak was aligned with the corresponding experimentally observed peak. Additional calculations were also performed, with the fully localized limit double counting scheme. This did not lead to significant differences, but only to a small relative shift between the $4f$ multiplets and the $[spd]$ density of states. These spectra are essentially similar to the spectra presented below, see Appendix~\ref{app:LDA_GGA_FP_FLLL}.

In Figs.~\ref{fig:spectra1} and \ref{fig:spectra2} the photoemission spectra are shown. In the left panel of each subfigure the X-ray photoemission spectroscopy (XPS) spectra are displayed and in the right panel the inverse photoemission spectroscopy, \textit{i.e.} Bremstrahlung isochromat spectroscopy (BIS) spectra. For a RE metal with $n$ $4f$-electrons, the XPS spectrum corresponds to  $f^n \rightarrow f^{n-1}$ transitions. The BIS spectrum corresponds to $f^n \rightarrow f^{n+1}$ transitions. For an easier comparison we displayed, apart form the bare spectral function, also the same spectral function but modified according to estimated cross-sections of the $d$ and $f$ states, with Gaussian smearing and with an estimation of the background. The cross-sections for the $d$ and $f$ orbitals, are estimated from their magnitude around the Fermi-level calculated with the single-scatterer final-state approximation~\cite{winter_cross, Redinger, Marksteiner_cross}. We used a Gaussian smearing of FWHM$=0.3$~eV for XPS and FWHM$=1.0$~eV for BIS to simulate the spectrometers resolution. We added a background proportional to the integral under the spectrum, to simulate the experimental background due to electrons (XPS) or photons (BIS) that scatter before they reach the detector. The XPS and BIS spectra were normalized such that the first occupied and first unoccupied peaks have approximately the same height as in experiment. There is an excellent agreement between theory and experiment as Figs.~\ref{fig:spectra1} and~\ref{fig:spectra2} show. All peaks in the atomic multiplet structure are captured. For some elements the different excitations are separated a bit too much compared to experiment. This separation is directly related to the values of $U$ and $J$. Since we did not tune $U$ to fit experiment, but kept it constant across the series, a small mismatch between calculation and observation may be expected. Also the Hund's $J$ seems to be a bit overestimated in general. 

The Ce BIS spectrum captures the multiplet features very well. The shoulder just above the Fermi level is not completely captured by the bare spectral function, however when taking into account the ratio between the $f$ and $d$ cross-sections just above the Fermi level the theoretical spectrum agrees better with experiments. In Ce however, effects involving hybridization might play also a role~\cite{held01prl87:276404,PhysRevLett.96.066402}. 
For the elements after Ce, the hump in the XPS spectrum just below the Fermi level (between 0 and -2 ~eV) is captured quite well, even by the bare spectral function and originates mainly from the $d$ electrons. This shoulder was not identified in the studies by Leb\`egue et al ~\cite{Lebegue2006_light,Lebegue2006_heavy}. Taking into account the $f$ and $d$ cross-sections just below the Fermi level, even improves the description of this feature, except for Sm. However for Sm it is suggested that this feature partially arises from divalent Sm atoms present at the surface~\cite{PhysRevLett.40.813,PhysRevLett.88.136102,PhysRevB.19.6615,PhysRevLett.63.187}
For Pr, Nd and Sm the relative height of the peaks in the BIS spectrum seems to be in better agreement with experiment than was found in Ref.~\cite{Lebegue2006_light}. Also the relative height in the XPS spectrum of Tb and Dy seems to be more accurate than found in Ref.~\cite{Lebegue2006_heavy}. There are three main differences between the present calculations and the calculations done in Refs.~\cite{Lebegue2006_light,Lebegue2006_heavy}. The first is that in both Refs.~\cite{Lebegue2006_light,Lebegue2006_heavy} the atomic sphere approximation (ASA) was used, whereas we use a full-potential code. The second difference is that in those works the crystal field splitting was not taken into account in the atomic impurity problem. The third difference might be that the temperature used in the Boltzman factors for the calculation of the one-electron Green's function, was probably different. 

In general we find that the HIA reproduces the experimental XPS and BIS spectra very well. This might be an indication that the full-potential treatment and non-sphericity of the potential experienced by the 4f orbitals is of significant importance. We finally want to mention that we have also performed a comparison between HIA and LDA+U for selected elements, e.g. Tb, in order to analyze how close these two approaches really are. The LDA+U results (see Appendix~\ref{app:LDAU}) show significant differences with HIA and with the available experimental data. In Ref.~\cite{peters14prb89_205109}, where the example of TbN was investigated with several methods, the authors already concluded that HIA was superior to LDA, and LDA+U, there were however no experimental data for the photoemission spectra available.



\section{Conclusion}
In this manuscript we have examined the applicability of the Hubbard I approximation (in connection with a full-potential electronic structure method) and how this method reproduces cohesive, structural and magnetic properties, as well as spectroscopic data, of the rare-earth series. We find good agreement between theory and observations, where a comparison can be made. In particular it is rewarding that equilibrium volumes, bulk moduli and magnetic properties are in good agreement with measured data. Similarly, calculated magnetic excitations as well as photoelectron spectra (direct and inverse) are in good agreement with measured data. As to the $4f$ magnetic moment we obtain similar values as would be obtained from a Russel-Saunders ground state. It is rewarding that this follows as a natural result from a quantum mechanical treatment that makes no assumption of the mechanism of coupling angular momentum states. We have also pointed to shortcomings of other methodologies, like LDA and LDA+U, in establishing results that consistently agree with measurements. In particular the electronic structure from these theories is found to not reproduce the measured XPS and BIS spectra, while the Hubbard I approximation gives a very satisfactory account of the measured spectra.

Among the different methods considered here, the treatment of the $4f$ shell as part of a non-hybridizing core comes closest to the Hubbard I approximation, since the LDA+U approximation is found to overestimate the hybridization and results in formation of dispersive energy states. We thus considered for simplicity exchange parameters of the rare-earth elements with this method. The $4f$ states were treated in the core, with a Russel-Saunders ground state. The resulting exchange parameters give ordering temperatures and magnon dispersion that are in acceptable agreement with measurements. The $4f$ induced polarization of the $[spd]$-valence band states is also captured with this $4f$ in the core treatment, a poor-mans version of the  Hubbard I approximation.

The Hubbard I approximation is hence demonstrated to be consistent with the standard model of the lanthanides, which identifies the $4f$ shell as atomic-like, and provides practical and reliable theoretical framework of the rare-earth elements and rare-earth containing materials in general. This opens for accurate theoretical analysis of rare-earth containing multiferroics, rare-earth based permanent magnets, rare-earth based topological insulators, and rare-earth based photovoltaics.

Although we think that the HIA is, among the  available state-of-the-art methods, the most promising for the REs, a correct assessment of the magnetic anisotropy and related quantities, remains a challenge. These quantities strongly depend on the subtle balance between the crystal field and the spin-orbit coupling. From the current work we understand the importance of some technicalities of the simulations. 
One important ingredient is the full-potential treatment. Another important component is the charge self-consistency of the simulations in order to correctly describe the valence electrons. Here, however, the methodological issue of the double counting correction arises. We also have to investigate, whether HIA properly describes the crystal field to the accuracy needed for these sensitive quantities. Similarly, in the future it would be interesting to address the cohesive properties in the magnetically ordered phases. We are positive that HIA is potentially able to provide an equally accurate picture for magnetic and non-magnetic phases, but methodological advances are needed.

\section{Acknowledgements}
The work of D. C. M. Rodrigues was funded by CAPES and CNPq (Brazil).
A. Bergman and O. Eriksson acknowledge the support from the Swedish strategic research programme  eSSENCE and Vetenskapsr\r{a}det (VR). O. Eriksson also acknowledges support from the Knut and Alice Wallenberg (KAW projects 2013.0020 and 2012.0031) foundation, projects 2013.0020 and 2012.0031. M. I. Katsnelson acknowledges support from European Research Council (ERC) Advanced Grant No. 338957 FEMTO/NANO. A. Delin acknowledges financial support from Vetenskapsr\r{a}det (VR), The Royal Swedish Academy of Sciences (KVA), the Knut and Alice Wallenberg Foundation (KAW), Carl Tryggers Stiftelse (CTS), Swedish Energy Agency (STEM), and Swedish Foundation for Strategic Research (SSF).
The authors acknowledge the computational resources provided by the Swedish National Infrastructure for Computing (SNIC), at Uppsala Multidisciplinary Center for Advanced Computational Science (UPPMAX), at the National Supercomputer Centre at Link\"oping University and at the PDC Center for High Performance Computing at the KTH Royal Institute of Technology. Moreover we would like to thank Patrik Thunström.



\appendix

\section{Importance of charge self-consistency in the HIA method\label{app:cscHIA}}
\begin{figure}[!h]  
\hspace{-0.8cm}
\includegraphics[angle=0,width=74mm]{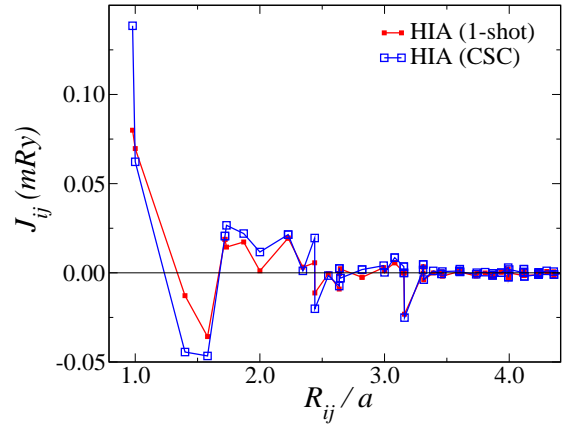} 
\caption{Effective exchange parameters in hcp Tb, extracted from single-shot and charge self-consistent HIA calculations. A positive sign corresponds to the FM coupling.}
\label{tb-hia-jij}
\end{figure}
We have performed a series of calculations for elemental Tb to check the differences in the calculated $J_{ij}$'s, obtained with single shot HIA and charge self-consistent HIA. In the latter, the updated electron density is used to construct a new Kohn-Sham potential. This is done in an iterative manner until the potential is converged.
For the spin-polarised HIA calculations we have neglected the effect of SOC and solved an impurity problem for the spin-polarised LDA (LSDA) solution. 
This recipe implies that the double counting potential is different for spin-up and spin-down electrons.
In the present case, we adopted FLL formulation, but in order to converge to the saturated magnetic solution, we had to significantly decrease the Hund's $J$ parameter (by factor two in case of Tb) with respect to Table~\ref{tab:J}.
This is due to the fact that $J$ is supposed to compensate the exchange splitting, present in the LSDA, since it will be reintroduced through the spin-polarised self-energy.
However, if the magnitude of this compensation is too large, it results into the flip of the spin moment. 
Thus, it starts to fluctuate from one iteration to another and the self-energy can never be converged.

The calculated exchange parameters obtained with single-shot and charge self-consistent HIA are shown in Fig.~\ref{tb-hia-jij}. The first two NN interactions are the most dependent on the description of the $4f$ states.
More distant neighbours, which are due to RKKY mechanism, are less affected by these details.

\begin{figure}[]  
\includegraphics[angle=0,width=0.8\columnwidth]{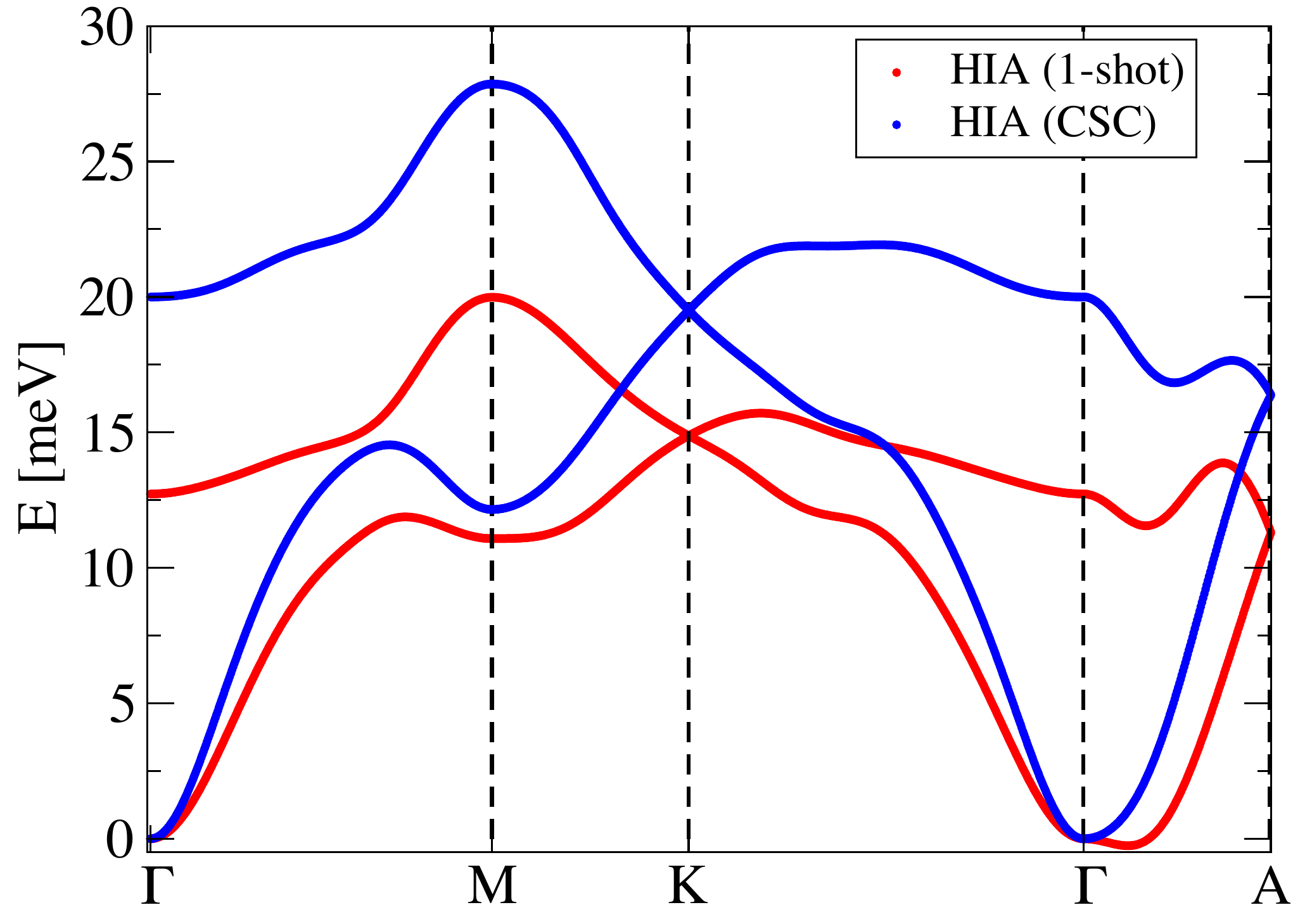} 
\caption{Calculated adiabatic magnon spectra of Tb using the $J_{ij}$-parameters taken from Figure~\ref{tb-hia-jij}.}
\label{tb-hia-magnon}
\end{figure}
Next we have used these $J_{ij}$-parameters to calculate the adiabatic magnon spectra. 
The results, shown in Figure~\ref{tb-hia-magnon}, indicate the importance of the charge self-consistency in the HIA calculation.
One-shot HIA calculations produce an unstable ferromagnetic state, which is indicated by the magnon instabilities specifically along $\Gamma$-A direction.  For the late REs we found a significant shift from $[spd]$ character to $f$ character in the LDA calculation when treating the $4f$ electrons as valence electrons, which is the starting point for the HIA. A single-shot HIA restores the correct $4f$ occupation, but does not allow for the remaining $[spd]$ electrons to relax to the adjusted potential. The effect of this relaxation, obtained in a charge self-consistent (CSC) calculation, is that the $[spd]$ density contracts towards the nucleus. The results in Figure~\ref{tb-hia-magnon} indicate that the $[spd]$ contraction is important for a proper description of the magnetic interaction.

\begin{figure}[]  
\vspace{-0.8cm}
\includegraphics[angle=0,width=\columnwidth]{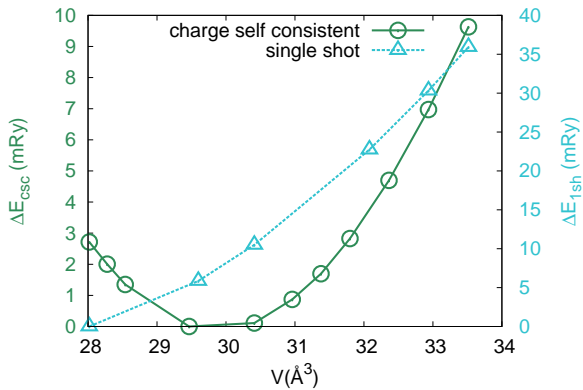}
\vspace{-1cm}
\caption{Energy versus volume curve for Gd calculated with single shot HIA (blue squares, right axis) and charge self-consistent HIA (green circles, left axis), using an LDA functional.}
\label{Gd-hia-1sh}
\end{figure}
The relaxation of the $[spd]$ electrons, obtained in a CSC calculation, also significantly affects the cohesive properties, such as the equilibrium volumes and bulk moduli, as is illustrated with Gd as example in Figure~\ref{Gd-hia-1sh}. As described above, the LDA starting point for the HIA calculation, overestimates the $f$ character at the expense of $[spd]$ character. The single-shot HIA calculation removes the $f$ electrons from the bonding and restores the correct $f$ occupation and thereby increases the $[spd]$ occupation. The first effect results in less binding, whereas the second effect results in more binding. For the late REs the second effect is usually stronger and single-shot HIA largely overestimates the binding and therefore underestimates the volume. Allowing the potential to adjust in a charge self-consistent way to the changes in the electron density, improves the description of the $[spd]$ electrons. The $[spd]$ electrons contract towards the nucleus, which results in less overlap with the wave functions on neighboring atoms. Hence the CSC procedure results in less binding and therefore a bigger volume than single-shot HIA. Since the $[spd]$ electrons are responsible for the chemical bond, a proper description of them is essential. For the volumes, we found it is especially important to use charge self-consistent simulations in the heavy REs. For the early REs the effects were less pronounced.\\

\begin{figure}[]
\begin{center}
     \subfigure[ Volume\label{subfig:vol_diffA}]{\includegraphics[width=0.8\columnwidth]{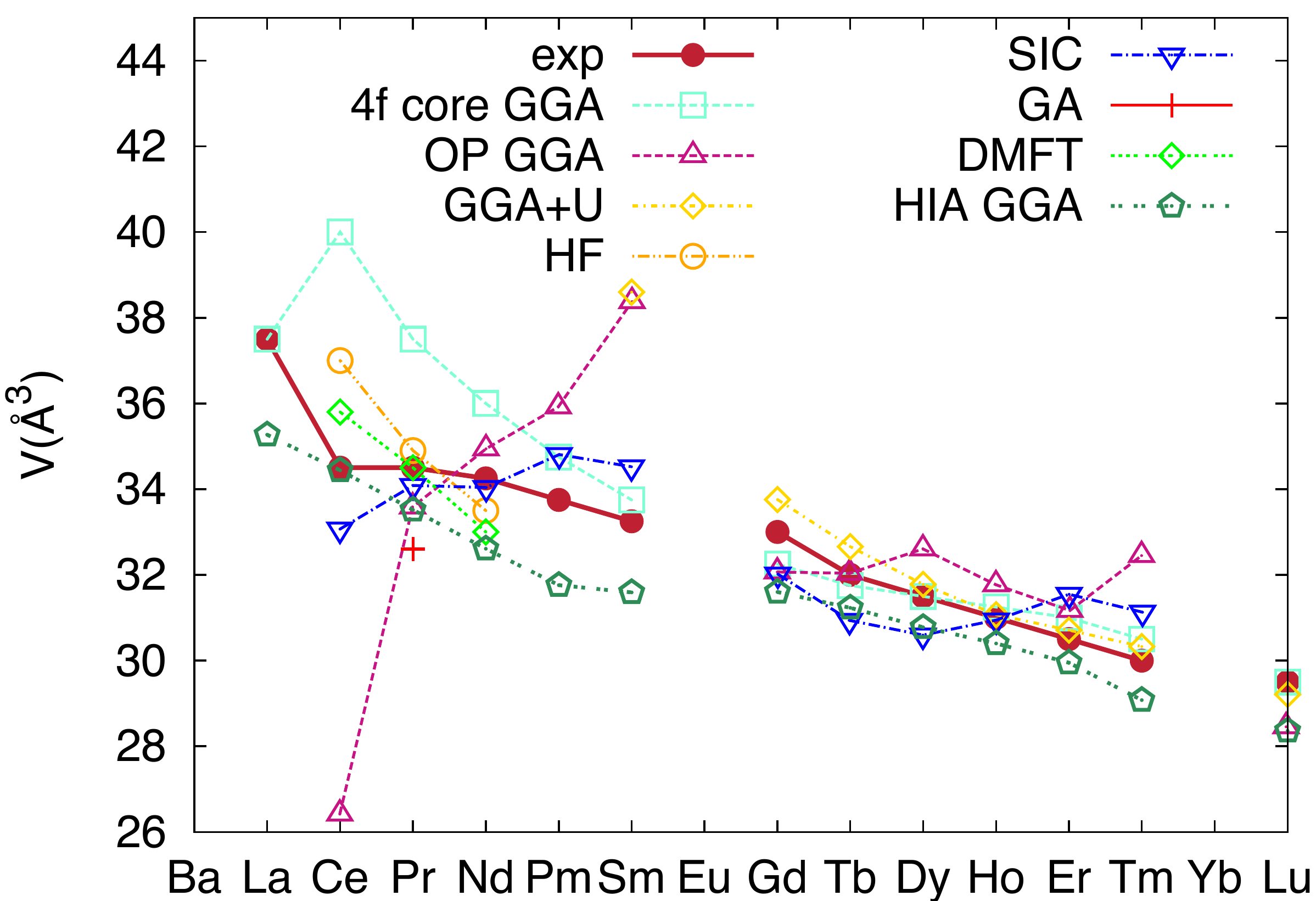}} 
     \subfigure[ Bulk modulus\label{subfig:bulk_diffA}]{\includegraphics[width=0.8\columnwidth]{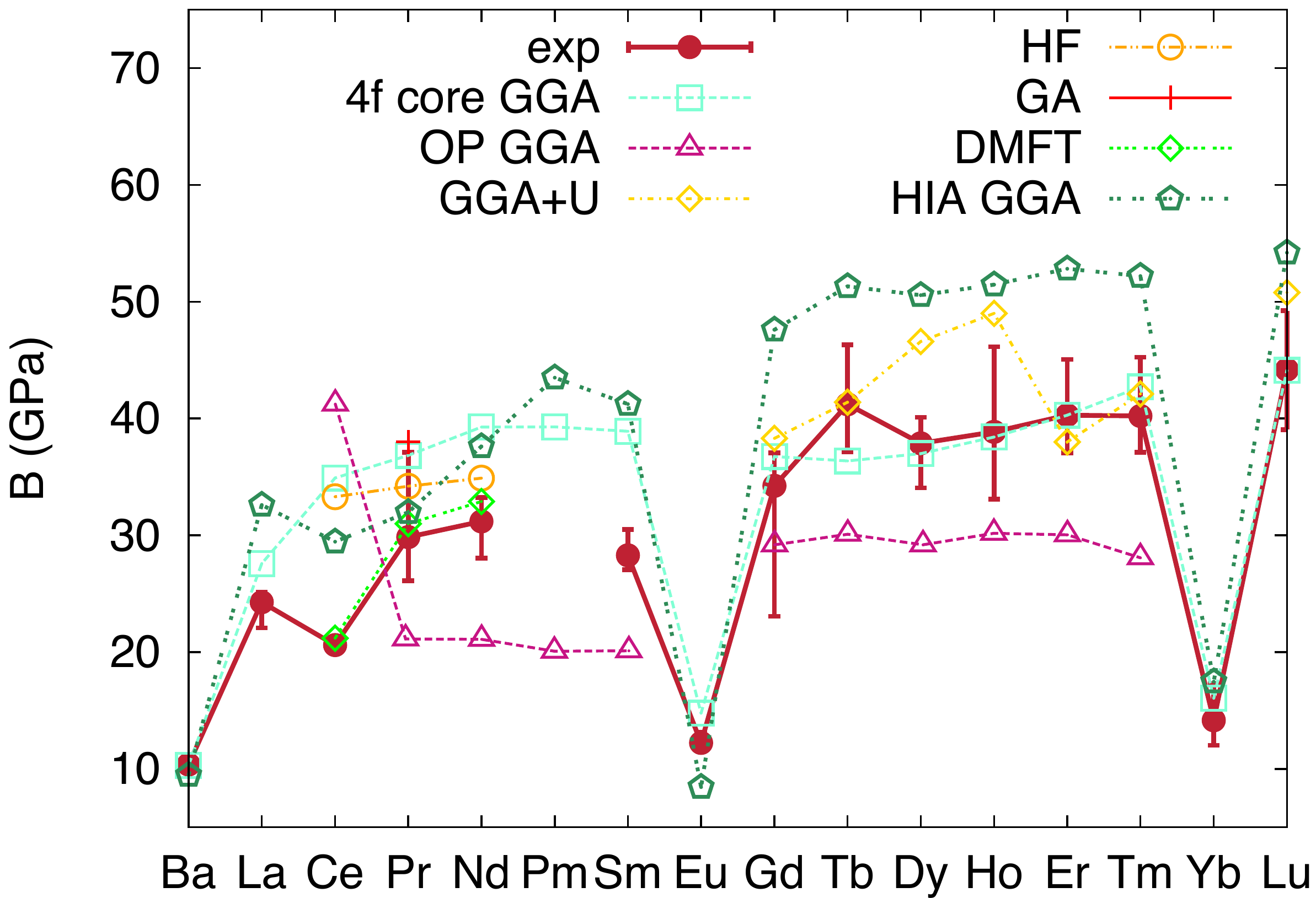}}   
\caption{Comparison between different methods for the volume (a) and bulk modulus (b) of the rare-earths. Experiments were taken from  K.A. Gschneidner~\cite{exp_V} and W. A. Grosshans {\it{et al.}}~\cite{exp_B}. Results calculated with $4f$-in-core GGA treatment were taken from A. Delin {\it{et al.}}~\cite{Delin1998}. The GGA + Spin orbit coupling with orbital polarization (OP GGA) results were taken from P. S\"oderlind {\it{et al.}}~\cite{soderlind}. The self interaction correction (SIC) results are from P. Strange {\it{et al.}}~\cite{PStrange} and the GGA+U results from S.K. Mohanta {\it{et al.}}~\cite{Mohanta20101789}. The Hartree-Fock (HF) and Dynamical Mean Field Theory (DMFT) results for Ce, Pr and Nd, were taken from A. K. McMahan~\cite{McMahanPRB2005}. The Gutzwiller approximation (GA) results for Pr were taken from N. Lanat\`{a} et al~\cite{Lanata_PhysRevX.5.011008}.}
\label{V_diffAuth}
\end{center}
\end{figure}

\section{Cohesive properties, different methods\label{app:compdiffmeth}}

In the main text we briefly compare the cohesive properties obtained with the Hubbard I method to the same properties obtained with other methods, such as the $4f$ core model, GGA+U, SIC and others. In Fig.~\ref{V_diffAuth}, we compare our results to those obtained by other groups~\cite{Delin1998,soderlind,PStrange,Mohanta20101789,McMahanPRB2005}. It is clear that it is most difficult to describe the beginning of the RE series. None of the methods captures the pronounced decrease in volume between La and Ce and the moderate decrease between Ce and Sm.

\begin{figure}[]     
     \subfigure[ HIA\label{subfig:HIA}]{\includegraphics[width=.235\textwidth]{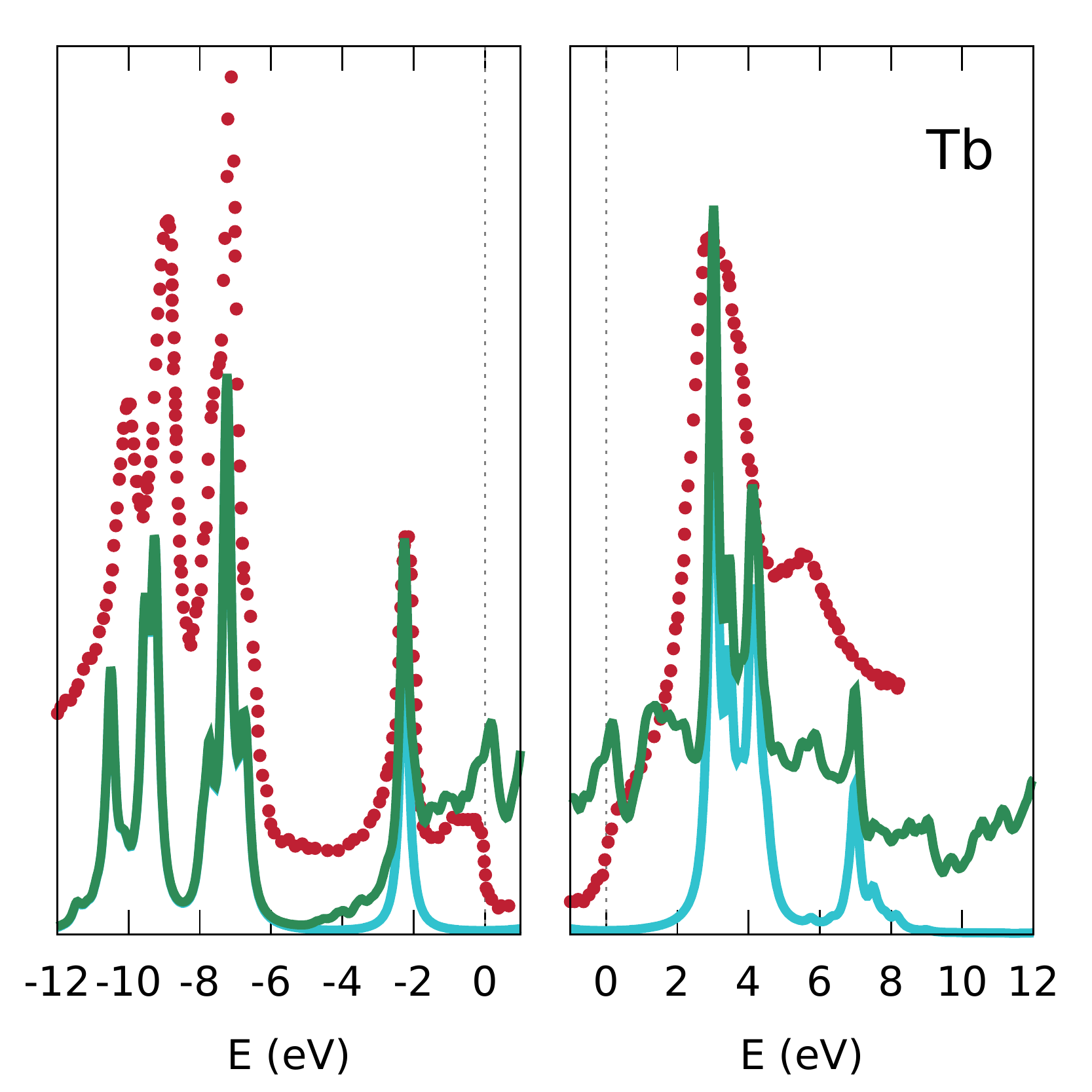}} 
     \subfigure[ LDA+U\label{subfig:LDA+U}]{\includegraphics[width=.235\textwidth]{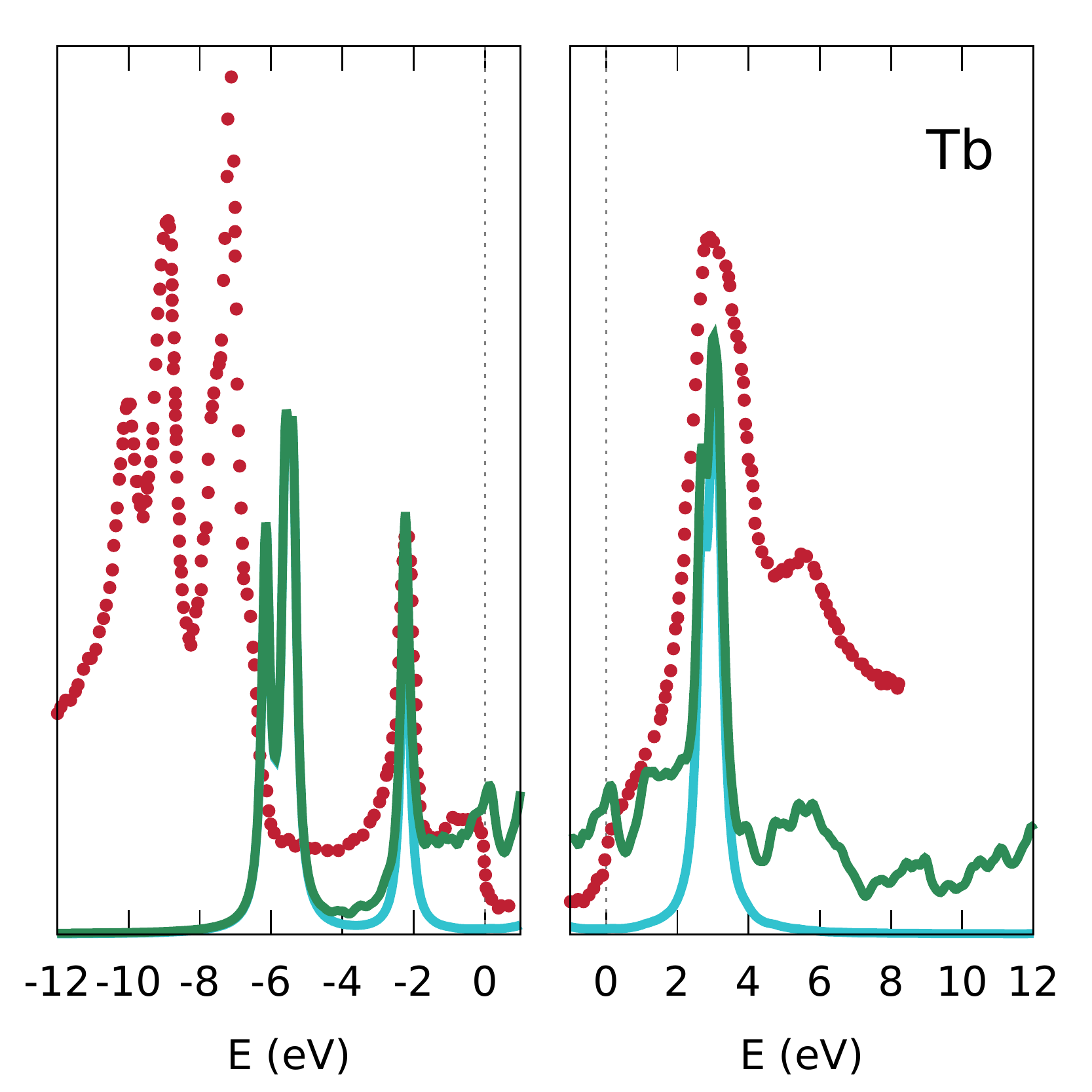}}  
\caption{Comparison between the spectra calculated with the HIA and with LDA+U. The red dots are experimental data from Ref. \cite{LangCox1981}. The green line is the calculated total spectral function. The blue line is the calculated $4f$ contribution to the spectral function. For each method the plot consists of the XPS part (left) and BIS part (right).}
\label{fig:TbLDA+U}
\end{figure}

\begin{figure*}[p]
     \centering
     \subfigure{\includegraphics[width=.32\textwidth]{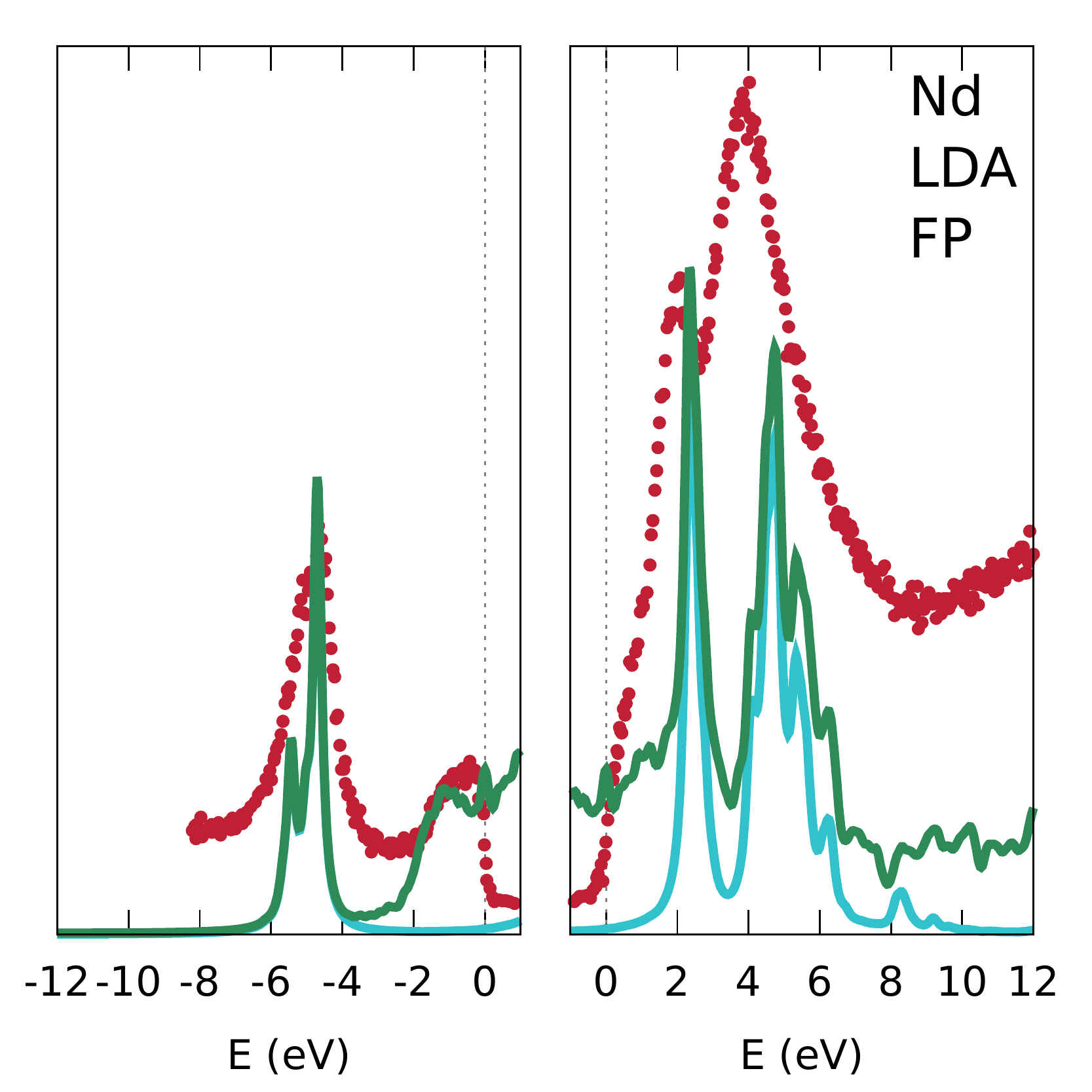}}
     \subfigure{\includegraphics[width=.32\textwidth]{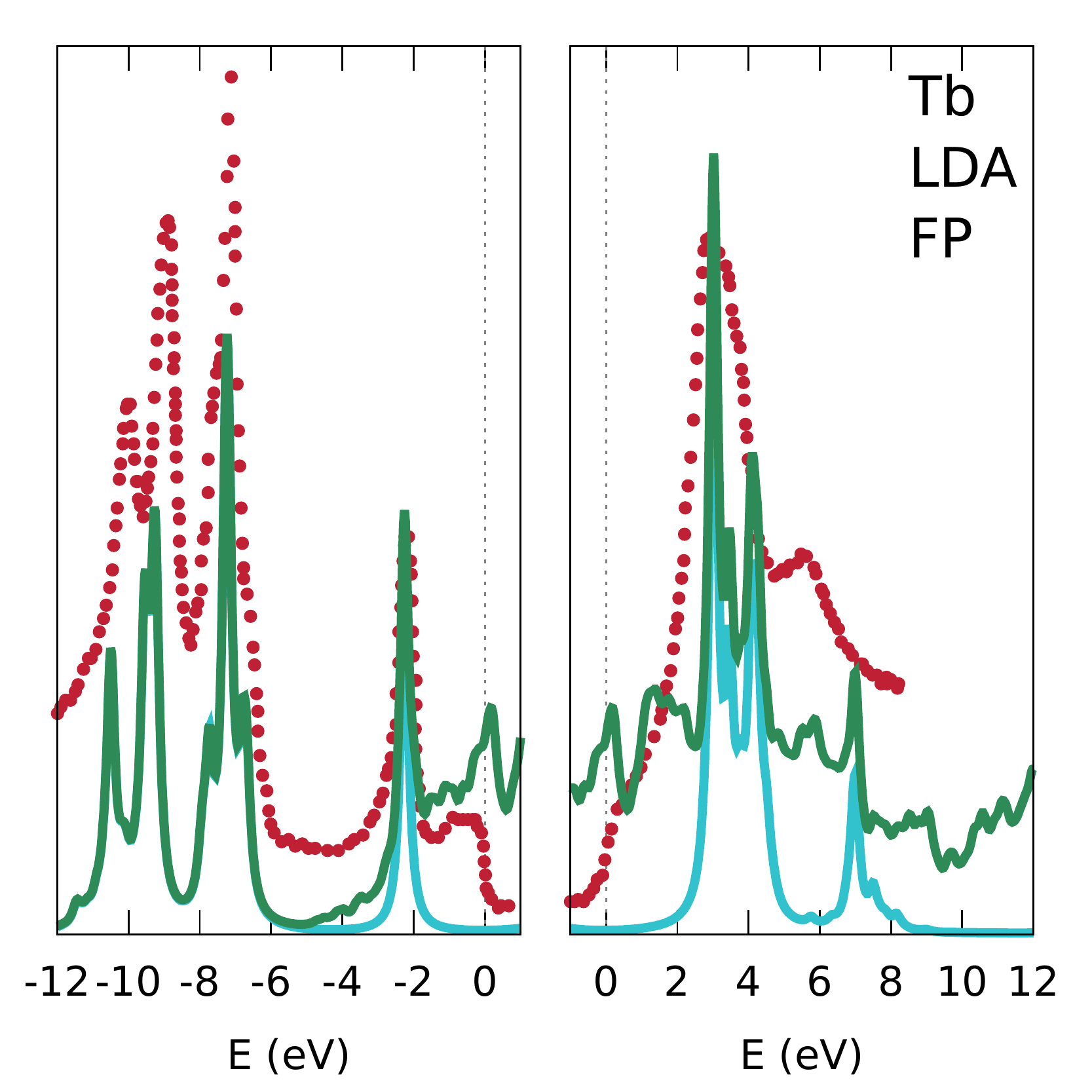}}
     \subfigure{\includegraphics[width=.32\textwidth]{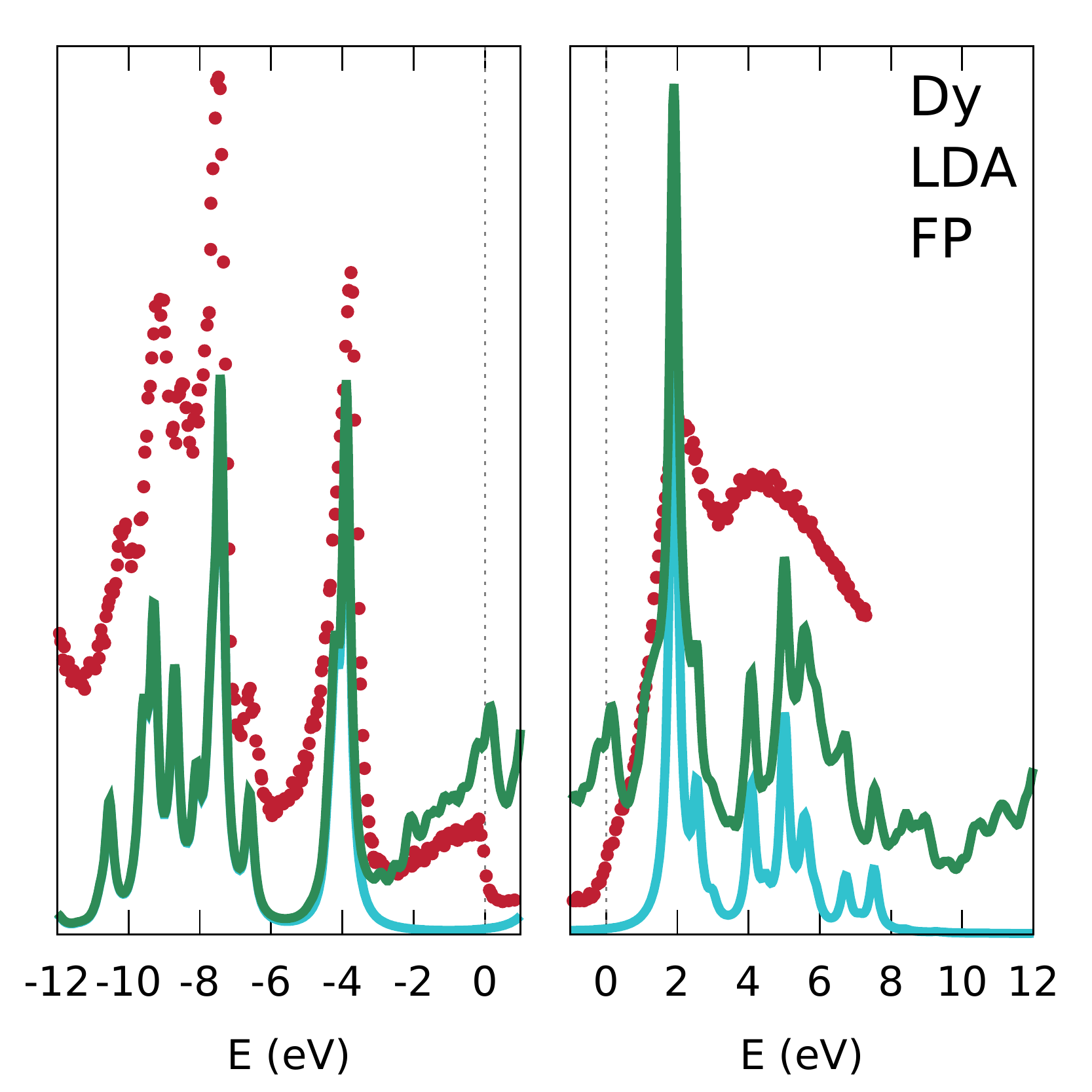}}
     \subfigure{\includegraphics[width=.32\textwidth]{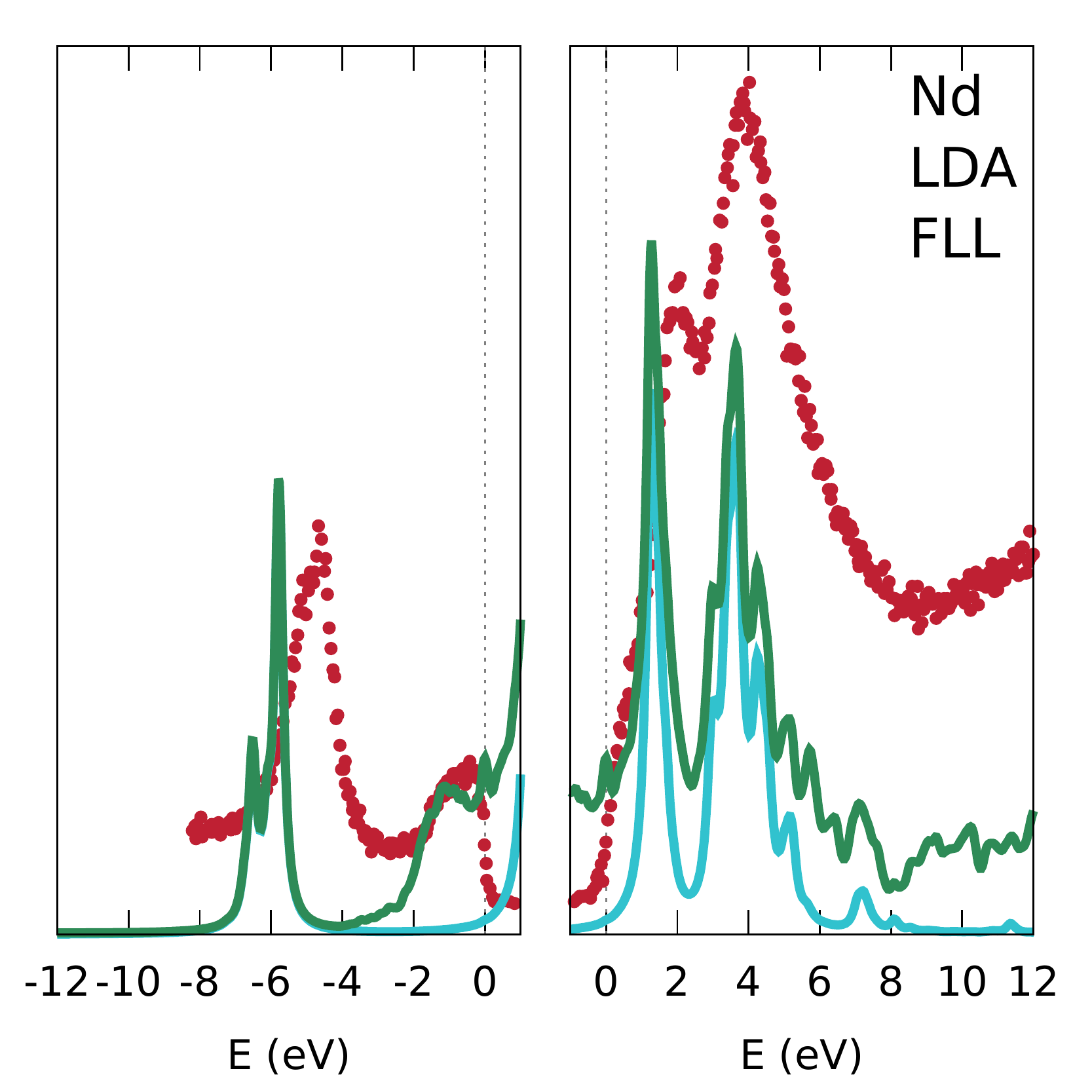}}
      \subfigure{\includegraphics[width=.32\textwidth]{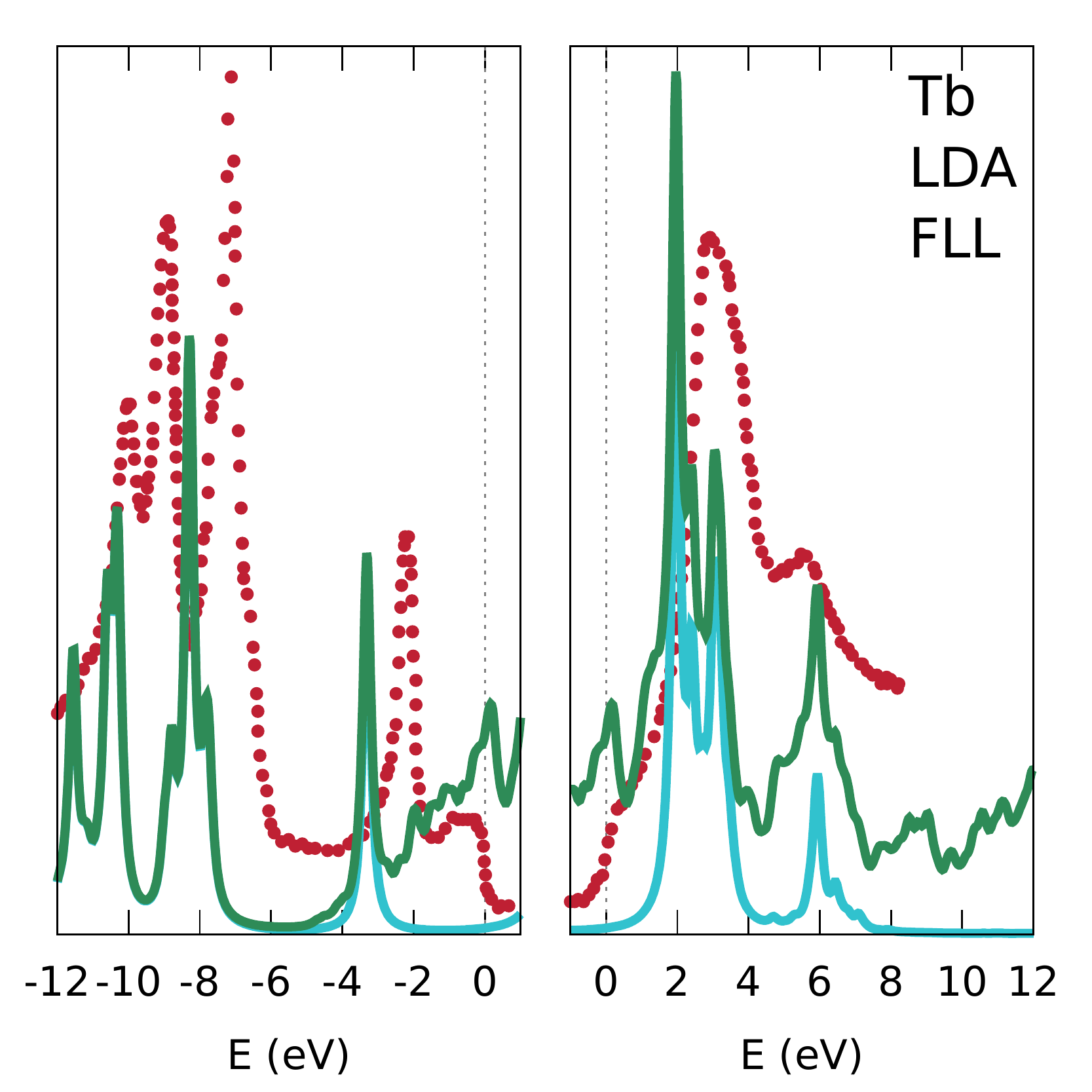}}
      \subfigure{\includegraphics[width=.32\textwidth]{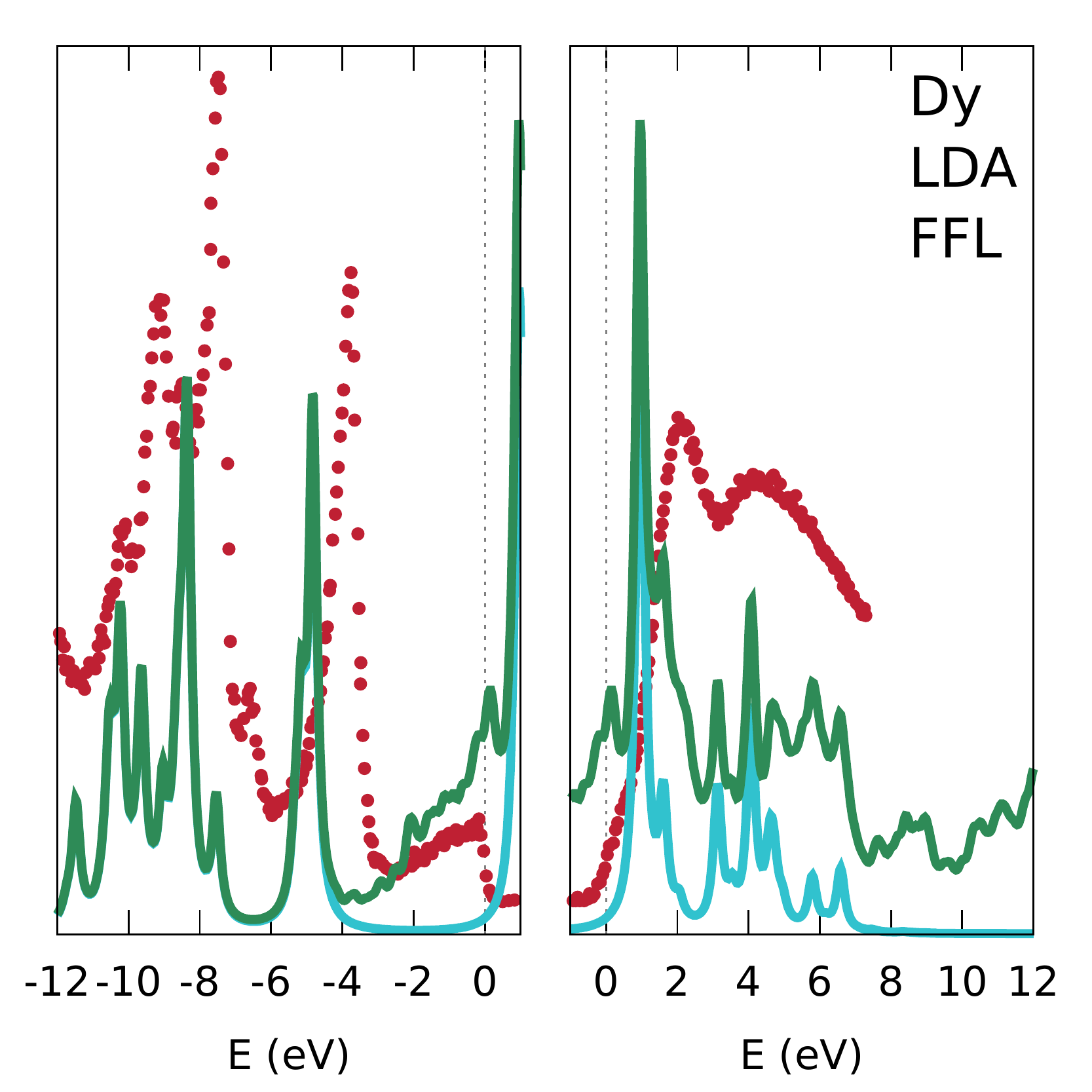}}
     \subfigure{\includegraphics[width=.32\textwidth]{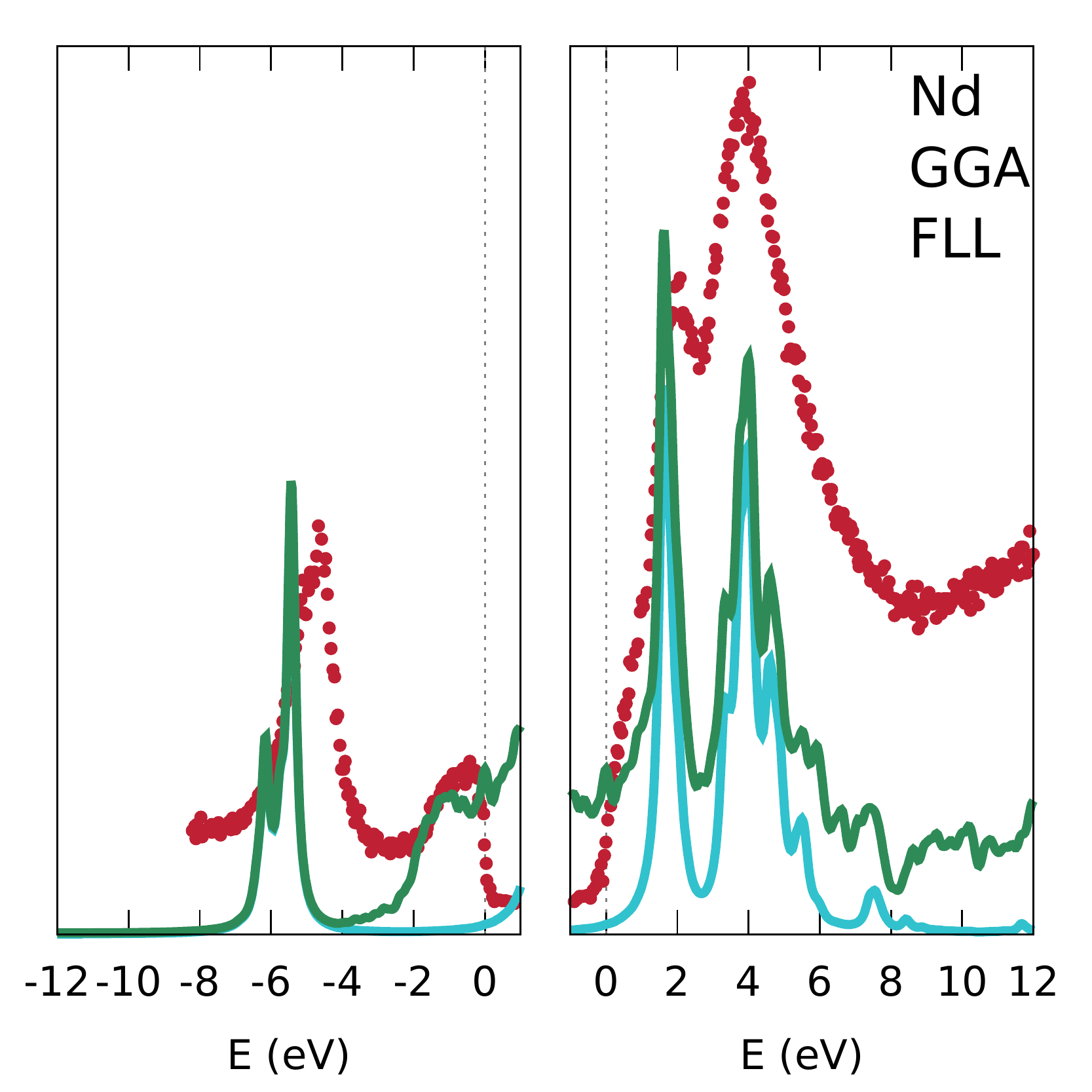}}
      \subfigure{\includegraphics[width=.32\textwidth]{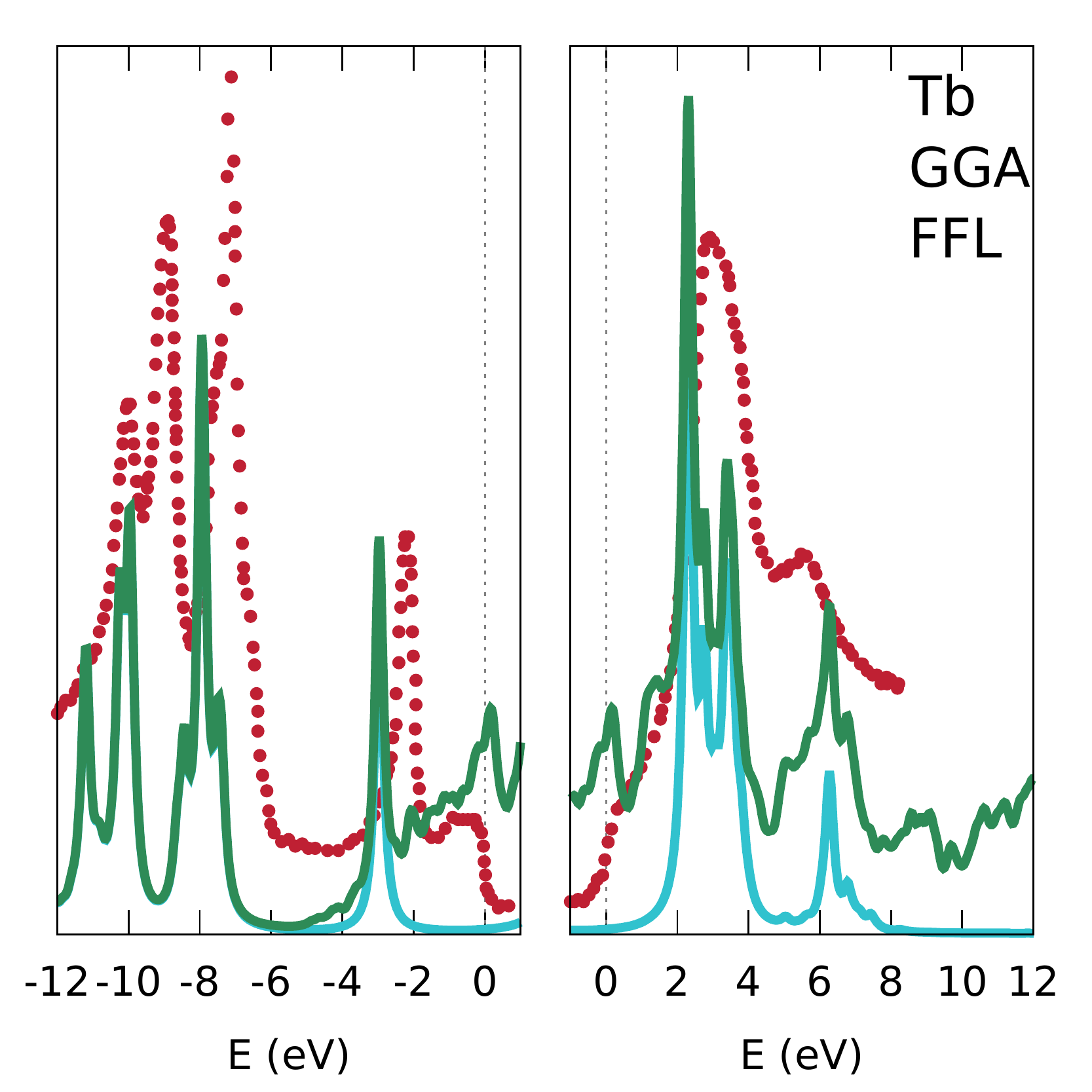}}
      \subfigure{\includegraphics[width=.32\textwidth]{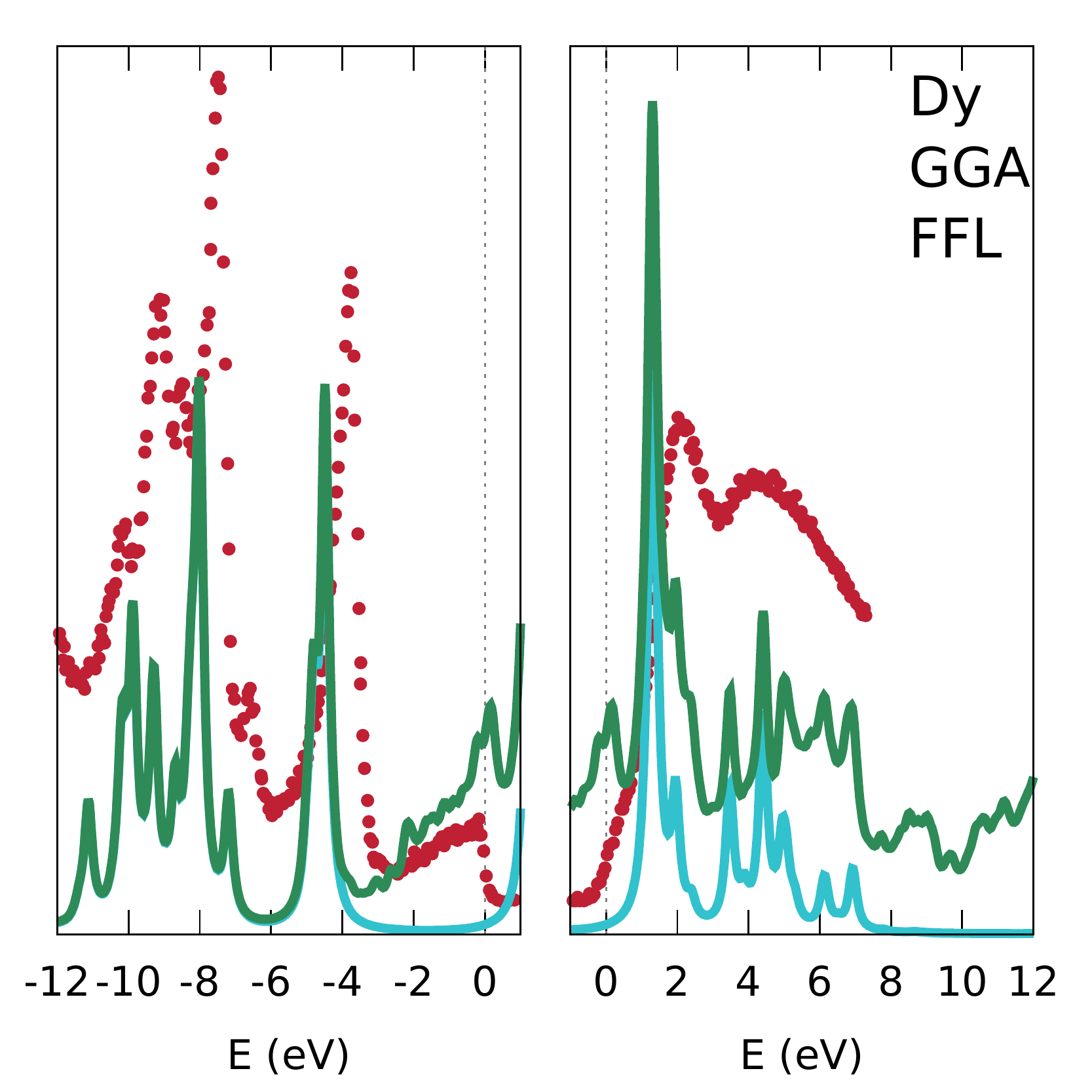}}
\caption{Spectra for Nd, Tb and Dy: The first and second row show the difference between the fixed peak (FP) double counting and the fully localized limit (FLL) double counting. The second and third row show the difference between LDA and GGA functionals. The red dots are experimental data from Ref.~\cite{LangCox1981}, the green line is the calculated total spectral function and the blue line is the calculated $4f$ contribution to the spectral function.  \label{fig:spectraLDAGGA_PFFLL}}
\end{figure*}

\section{Photoemission spectra with different exchange correlation functionals and different double countings\label{app:LDA_GGA_FP_FLLL}}
For the calculation of the spectra, we use a double counting scheme that fixes the position of the first occupied or unoccupied peak, with the only purpose to make comparison with experiment easier. In the top two rows of Fig.~\ref{fig:spectraLDAGGA_PFFLL} we compare this double counting scheme to the Fully Localised Limit double counting scheme, as defined in Eq.~\ref{eq:DCFLL}, for three different elements. The main difference between these two schemes is a rigid shift of the $f$ spectrum with respect to the $[spd]$ spectrum.

In Sec.~\ref{sec:spectra} we also mention that changing from an LDA to a GGA functional does not significantly change the spectra. In the bottom two rows of Fig.~\ref{fig:spectraLDAGGA_PFFLL}, we compare the spectra calculated with two different functionals, for three elements. The spectra are essentially the same.

\section{Photoemission spectra: LDA+U versus LDA+HIA \label{app:LDAU}}

The Hubbard I approximation is very suitable to describe the lanthanides and outperforms the LDA+U method. In Figure~\ref{fig:TbLDA+U}, we compare the photoemission spectra calculated with HIA and LDA+U for Tb.  The HIA spectrum (Figure~\ref{subfig:HIA}) captures all multiplet features and is clearly more accurate than the LDA+U spectrum (Figure~\ref{subfig:LDA+U}). Although LDA+U can describe several properties quite well, it is essentially a single particle method and can not be expected to reproduce many-body multiplet features accurately.

The setup for the HIA calculation was described in Sec. \ref{sec:spectra}. The same Hubbard $U=7$~eV and Hund's $J=1.080$~eV was also used for the LDA+U calculation. For a fair comparison the same double counting scheme was used, where the double counting correction was chosen such that the first occupied peak is aligned between theory and experiment.

\clearpage
%

\end{document}